\documentclass[headings = small,
               fontsize = 11pt,
			   USenglish,
			   parskip = half,
			   oneside
			   ]%
               {scrartcl}

\usepackage[T1]{fontenc}
\usepackage[latin1, utf8]{inputenc}
\usepackage[USenglish]{babel}
\usepackage{csquotes}
\usepackage{bbm}
\usepackage{lmodern, microtype}
\usepackage{scrlayer-scrpage}
\usepackage{booktabs, longtable, tabularx, enumerate, array, float}
\usepackage{amsmath, amssymb, amsthm, mathtools, nicefrac}
\allowdisplaybreaks
\usepackage{graphicx, colortbl, xcolor, color}
\usepackage[doublespacing]{setspace}
\usepackage{lscape}
\usepackage{comment}
\usepackage{rotating}
\usepackage{multirow}
\usepackage{url}
\usepackage{pdflscape}
\usepackage{abstract}
\usepackage{multicol}
\usepackage{listings}

\usepackage[%                           
           round,                      
           semicolon,                  
           nonamebreak,                 
           authoryear
          ]%
          {natbib}                     
\clearmainofpairofpagestyles
\theoremstyle{plain}
\newtheorem{defi}{Definition}

\begin{document}
\begin{center}
\onehalfspacing 
\huge
Distributional Reference Class Forecasting of Corporate Sales Growth With Multiple Reference Variables
\footnote{The author thanks Dominik Wied for valuable discussion and advice on this paper and Daniel Ziggel for bringing the topic of reference classes to his attention.
Further, the author is grateful to colleagues at Flossbach von Storch AG for providing the analysts' estimates. 
Moreover, the author thanks the Regional Computing Center of the University of Cologne for providing computing time on the DFG-funded  High Performance Computing system CHEOPS (Funding number: INST 216/512/1FUGG).
Wharton Research Data Services (WRDS) was used in preparing this paper.
This service and the data available thereon constitute valuable intellectual property and trade secrets of WRDS and/or its third-party suppliers. 
This research did not receive any specific grant from funding agencies in the public, commercial, or not-for-profit sectors.}
\end{center}
%\begin{center} \Large This version: \today \end{center}

\onehalfspacing
\begin{center} Etienne Theising
\\
\textit{Institute of Econometrics and Statistics\\ University of Cologne}\\
May 6, 2024
\end{center}

\begin{abstract}
This paper introduces an approach to reference class selection in distributional forecasting with an application to corporate sales growth rates using several co-variates as reference variables, that are implicit predictors.
The method can be used to detect expert or model-based forecasts exposed to (behavioral) bias or to forecast distributions with reference classes.
These are sets of similar entities, here firms, and rank based algorithms for their selection are proposed, including an optional preprocessing data dimension reduction via principal components analysis.
Forecasts are optimal if they match the underlying distribution as closely as possible.
Probability integral transform values rank the forecast capability of different reference variable sets and algorithms in a backtest on a data set of 21,808 US firms over the time period 1950 – 2019.
In particular, algorithms on dimension reduced variables perform well using contemporaneous balance sheet and financial market parameters along with past sales growth rates and past operating margins changes.
Comparisions of actual analysts' estimates to distributional forecasts and of historic distributional forecasts to realized sales growth illustrate the practical use of the method.
\end{abstract}


\newpage

\onehalfspacing
\section{Introduction}
A major aspect of statistics is to make projections and forecasts of future events which should be probabilistic in nature to reduce uncertainty \citep{dawid1984}.
To this end, we extend the method for distributional forecasts with reference classes proposed by \cite{theisingetal2023} to allow construction based on several co-variates.
Corporate bankruptcy, stock returns or cash flow items, e.g., are pivotal features in fundamental analysis of firms, key drivers of stock selection models \citep{gmx15} and in general challenging forecast tasks where long forecast horizons increase the level of complexity.
Particularly growth (rates) of corporate sales suffer from low predictability \citep{chankarceskilakonishok03}.
Forecasts thereof are often based on heuristics and were empirically shown to be biased as well as overoptimistic \citep[see, e.g.,][]{tvka1973, tvka1974, katv1973, cwd1988, dubu18}.

Forecast distortion is predominantly due to the \textit{inside view}, which considers each forecasting challenge as unique and neglects statistical information as well as results of similar forecast challenges \citep{kalo1993}. 
Hence, an inclusion of the \textit{outside view} in form of empirical data and existing experience can help to identify and reduce the aforementioned biases \citep{tega2016}.
The concept of the outside view in this scenario is the definition of a \textit{reference class} consisting of firms similar to the initial firm whose sales growth we want to predict \citep{kahnemantversky79,loka2003}. 
By means of this objective data subset of similar firms the forecaster is equipped to challenge and improve their forecast \citep{kahnemantversky79}.
Such adjustements of model based forecasts by experts \citep{woflo90,sari01,brufra17} or combinations of statistical forecasts with analysts' predictions \citep{lobo91,buwri91} are already established in the financial and forecasting literature.
Additionally, the resulting reference classes can be used to issue a distributional forecast and interval or point forecasts.

In general, constructing a reference class of comparable objects is known as the reference class problem in statistics.
While forecasting probabilities with respect to a given object, \citet{venn1888} notes that each object has several characteristics to determine a set of similar objects from which to derive these probabilities.
\citet{reichenbach1949} first called such a set a reference class.
Here, we consider firms and can imagine plenty of possible reference classes, e.g., all US firms, all S\&P500 firms, or firms with similar cash flow or stock market metrics.
Thus, we are challenged to find a reference class that is best in the sense of forecasting the distribution of, say, three-year sales growth.\footnote{\citet{venn1888} originally describes an example of a fifty-year-old consumptive Englishman with many possible reference classes, e.g., all humans, all males, all at least fifty-year-old Englishmen or all consumptive patients, that could be used for a distributional forecast of, e.g., remaining life expectancy.}
Clearly, the search for a reference class is paired with a specific forecast challenge in mind and a \textit{good} reference class always depends on this forecast challenge.

Reference classes and outside views are established in forecasting literature and practice, e.g., \cite{armstrong05} and \citep{tega2016} recommend the use of base rates, i.e. distributional information, known to improve forecasting performances \citep{ccmt16,karetal21}.
But in general the literature focused more on biases than on debiasing itself \citep{ccmt16}.
\cite{theisingetal2023} discuss some examples of reference class selection in the literature.
In our context, \citet{maubcal15} cover corporate sales growth and base reference classes on the current sales level, however, do not provide a theoretical justification or empirical tests of the procedure. 
In contrast, \cite{theisingetal2023} backtest reference class construction for corporate sales growth with a single co-variate and we generalize their method using rank based algorithms in order to handle several co-variates.

We propose a framework for reference class selection based on several co-variates, here called reference variables, and examine several approaches to find appropriate outside views to forecast corporate sales growth. 
Hence, we define reference classes for each company separately by means of additional observations that share similarities to the company at hand with respect to the reference variables. 
These approaches are easy to implement and we choose interpretable algorithms to build the reference classes. 
Thus, the proposed methods are well suited for practical application, the more so, as the outside view is straightforwardly provided by the realized sales growth rates within the reference class.
We recommend distributional forecasts in terms of the empirical cumulative distribution function (ECDF) of the reference class outcomes.
ECDFs are easy to calculate, non-parametric and include no assumptions on the underlying distribution.
Their simple structure empowers practitioners to investigate the reference class for a given company and discuss the nature of forecasts highlighting the procedure's interpretability.
Alternatively, a parametric model for sales growth distributions is discussed in \cite{stanleyetal96}.

The forecast performance of different algorithms and reference variables is backtested on the same data set as \cite{theisingetal2023} which consists of 21,808 US firms over the time period 1950 - 2019.
Probability integral transform values serve as a ranking of calibration.
This analysis yields that in particular past operating margins and past sales growth rates are suitable variables for reference class building regarding future sales growth rates and a subsequent distributional forecast thereof.
Dimension reduction using principal component analysis allows using more variables, e.g. contemporaneous balance sheet and financial market parameters, and shorter lags of past variables while simultaneously improving the results substantially by between 38\% and 71\%, depending on the forecast horizon.
Further, a case study compares the distributional forecasts with actual analysts' estimates, thus, illustrates the practical application of reference classes and how to apply their results in practice.
We additionally display historic distributional forecasts of sales growth rates and compare them to realized sales growth.

The remainder of the paper is organized as follows:
Section 2 contains the theoretical framework of reference class selection and the proposed algorithms.
Performance measures of reference variables and algorithms are covered in Section 3.
Further, Section 4 describes the data set used in the backtest presented in Section 5 along with the variable and model selection procedure.
Illustrative practical applications are demonstrated in Section~6.
Section 7 concludes and gives an outlook on future research.

\section{Reference Class Selection}

The notion of reference class forecasting traces back to theories of planning and decision-making under uncertainties and is motivated by the fact that forecasts are often based on heuristics and were empirically shown to be biased as well as overoptimistic \citep{kahnemantversky79}. 
A reason thereof is that forecasters and decision makers often focus solely on information on the specific case at hand, the inside view, while neglecting information on a class of similar cases, the outside view.
Statistical or empirical distributional information as well as base rates may serve as an outside view and can be seen as a data driven method to overcome overoptimism, wishful thinking or strategic misrepresentations.
\cite{kahnemantversky79} propose a corrective procedure for prediction, from the selection of a set of similar cases as a reference class to provide information on outcomes, over evaluation or estimation of the distribution of outcomes within the reference class and, finally, to a correction of an expert's inside forecast.
For the latter, they suggest an assessment of the expert's forecast predictability, e.g. the correlation between their predictions and the outcomes in case of linear prediction, and, e.g., a mean forecast is adjusted towards the mean of the reference class weighted by the forecast predictability. 

Here, we concentrate on a framework to select an appropriate reference class. 
This is critical as \cite{kahnemantversky79} gave no instructions how to build reference classes except the general rule to use similar cases. 
Moreover, there is a conflict of competing objectives in defining the reference class. 
On the one hand, taking as many cases into account as possible provides the most distributional information. 
However, it is crucial that each object is still comparable to the initial one and heterogeneity is limited. 
On the other hand, if the reference class only consists of elements extraordinarily similar to the initial object, the risk of an undersized and little informative reference class producing a likewise biased forecast exists.
Based on this trade-off \cite{loka2003} state: \textit{``Identifying the right reference class involves both art and science.''}

In literature, there are several studies dealing with reference class selection \citep[discusses a selection]{theisingetal2023} and a first concept with respect to the forecasting of future cash flows is proposed by \citet{maubcal15}. 
They state that sales growth is the most important driver of corporate value and define reference classes for sales growth by firms' real sales based on historical data of the S\&{}P1500 from 1994-2014 but their method has certain drawbacks \citep[see][]{theisingetal2023} and their study lacks a theoretical justification and an empirical test.
\cite{theisingetal2023} add a systematic analysis to the literature and evaluate forecast quality by backtesting the concept of \citet{maubcal15} along with refined reference class construction but the analysis is limited to using different but only a single reference variable.

In order to overcome this drawback we propose and systematically backtest different rank based algorithms that allow using multiple reference variables including an optional dimension reduction.
The approaches are easy to implement and interpret and find reference classes for each analyzed company separately.
Then, an assessment of the distribution within the reference classes follows directly from the outcomes within the reference class in shape of their ECDF and can be viewed as a distributional or probabilistic forecast.
This study evaluates the resulting reference classes by backtesting out-of-sample on a 1950--2019 data set to make a meaningful quality valuation. 
The following two subsections provide an extended theoretical foundation of \cite{theisingetal2023} and the proposed algorithms to select reference classes.

\subsection{Theoretical Framework}\label{subsec:theory}

For a given firm $i$ at time $t$, we aim to construct a reference class for an $h$-step ahead distributional forecast of sales growth $Y_{i,t}$.\footnote{We phrase the theoretical framework with a specific application in mind, namely forecasting corporate sales growth. For a general purpose the term `firm' can be replaced by `object' and the term `sales growth' can be replaced by `some characteristic'.} 
Any choice of reference class produces a forecast of the distribution of $Y_{i,t+h}$, that is a distributional $h$-step ahead forecast of the random variable $Y_{i,t}$ for individual $i$ at time $t$.
To this end, we assume that a sufficient amount of historical data on firms is available to assess the distribution of $Y_{i,t+h}$.

\begin{figure}
	\centering
	\includegraphics[width=.95\textwidth]{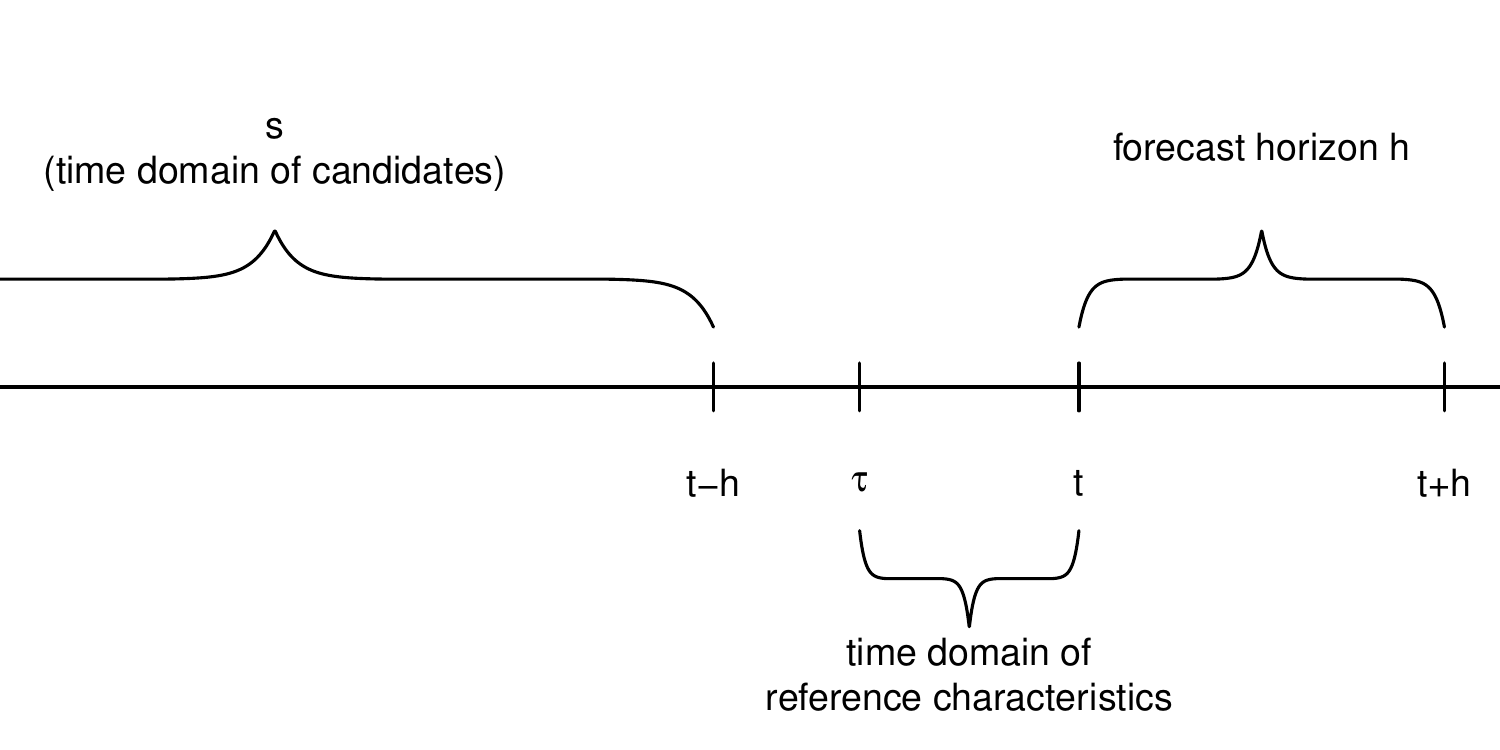}
	\caption{Illustration of reference class and prediction timeline. Firms $j$ at times $s$ denote the set of potential members (candidates) for the reference class of individual $i$ at time $t$. Note, $\tau \leq t-h$ is possible as well if $\tau \geq h$.}
	\label{fig:reference_scheme}
\end{figure}

We base the reference class on reference variables $X_{i, \tau :t} := \{ X_{i, t'} \}_{t'=\tau ,\dots ,t}$ and build a reference class $R$ by finding firms $j$ in the past which are similar to individual $i$ at time $t$ with respect to the reference variables.\footnote{The reference variables are also called reference characteristics or predictor variables in \cite{kahnemantversky79} and \cite{theisingetal2023} but we stick to the term `reference variable' as they are random variables here, used for reference class selection and do not predict directly but only implicitly throguh the selection.}
Similarity can be measured in multiple ways, for mathematical purposes it is convenient to view similarity according to some distance measure $d : D^2 \to [0, \infty )$. 
Then, $d( X_{i, \tau :t}, X_{j, \zeta :s} )$ shall be \textit{small}, where $s + h\leq t$ ensures that the realization of $Y_{j,s+h}$ is available and $D$ is the domain of $X_{i, \tau :t}$ (c.f. Figure \ref{fig:reference_scheme}).
$d$ could be a metric, e.g., based on some norm.
Thereby, we aim at finding neighbors for each firm separately.
A non-parametric forecast for the distribution of $Y_{i,t+h}$ is now given by the empirical cumulative distribution function of the values $Y_{j,s+h}$, $(j,s)\in R$ and serves as an outside view. 

Assumptions regarding the dependence of $Y_{i,t}$ on $X_{i, \tau :t}$ and the stability of this dependence over time are necessary \citep[c.f.][]{theisingetal2023}.
The first assumption behind the approach is the existence of a data generating mechanism, say a smooth function $f_h$ such that $Y_{i,t+h}\sim f_h(X_{i, \tau :t})$ which can be interpreted as the conditional distribution $Y_{i,t+h} | X_{i, \tau :t}$.\footnote{In case of a finance application like here, such a data generating mechanism may be called \textit{market mechanism}.}
Moreover, we need an assumption that this mechanism works similarly over time and we have $Y_{j,s+h} \sim f_h(X_{j, \zeta :s})$, $(j,s)\in R$, for the outcomes within the reference class, which can be interpreted as a kind of stationarity assumption.
If $X_{i, \tau :t}$ is close to $X_{j, \zeta :s}$, which is supposed to be provided by finding suitable reference classes, then $f_h(X_{i, \tau :t})$ is close to $f_h(X_{j, \zeta :s})$ and the empirical distribution function of $Y_{j,s+h}$ is a good approximation for the distribution of $Y_{i,t+h}$.
Note, the target is not to identify $f_h$, but to get information about reference adequacy.

\subsection{Proposed Algorithms for Reference Class Selection}

Algorithms for constructing a reference class from a given sample need to implement the aforementioned assumption regarding the stable dependence of sales growth $Y_{i,t}$ on reference variables $X_{i, \tau :t}$ and need to decide which past firm observations are similar enough with respect to the reference variables.
A window length parameter $w$ common to all algorithms defines the number of past years to use for a specific forecast challenge. 
$w$ selects observations from the sample to constitute a set of candidates $C$ for the reference class from a limited time period and thereby accounts for the degree of stability regarding the dependence.
Assessing the similarity to the firm of interest and deciding whether it is part of the reference class or not is the essential feature of each algorithm.

The decision problem of labeling each candidate `belonging to reference class' and `not belonging to reference class' makes the reference class selection a binary classification.
The selection is based on available co-variates (reference variables) only and not on the outcome of the candidate firms because these outcomes are not observed for the initial firm at hand as we seek to forecast this quantity.
Thus, we use unsupervised learning techniques to find firms that are sufficiently similar to the initial firm by algorithms.
The decision on sufficient similarity is part of the reference class problem, too, as it is an integral part of choosing a subset of candidates to form the reference class.

\begin{figure}
	\centering
	\includegraphics[width=.95\textwidth]{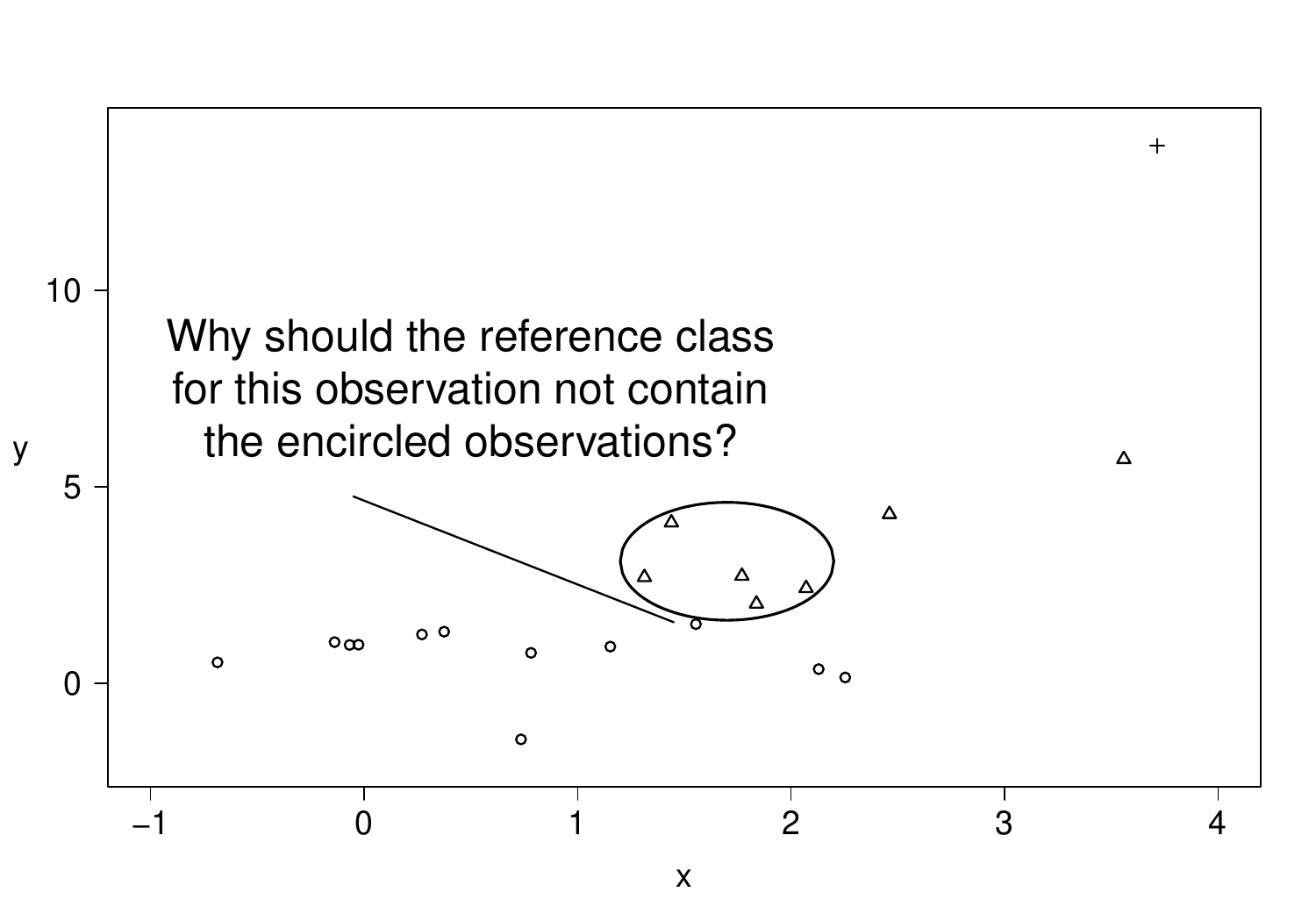}
	\caption{These three clusters constructed by the k-means algorithm for a simulated data cloud highlight the pitfall that elements on the border of one cluster may be closer to the elements of another cluster than to the majority of elements in their own clusters \citep[see][]{theisingetal2023}.}
	\label{fig:cluster_counterexample}
\end{figure}

In the application on firms we encounter skewed reference variables including outliers and use rank based methods to be robust against these data features.
Unsupervised cluster algorithms share the property to split the set of candidates in a fixed number of clusters and due to continuity of reference variables in our case we argue that it does not necessarily make sense to find candidates that are closest to the given firm in this manner (c.f. Figure \ref{fig:cluster_counterexample})\footnote{This applies to the special case here.
The proposed procedures do not account for `natural' clusters that might occur. 
For example, there might be only 20 observations `very similar' to the initial case.
But the algorithm may choose the most similar 25 observations and, thus, five less informative observations.
Hence, there is no general conclusion and the algorithms must be adapted to the specific forecast challenge.}.

We proceed with the following definition that a reference class has to consist of at least 20 elements or members in order to allow reasonable distributional forecasts.

\newpage
\begin{defi}
Reference Class\\
Let $C$ be a set of reference class candidates.
We call any set $R \subset C$ of observations $j$ at times $s$ reference class if the reference variables $X_{j, \zeta :s} = \{ X_{j, s'} \}_{s'=\zeta ,\dots ,s}$ and the outcomes $Y_{j,s+h}$ are observed and $\vert R\vert \geq 20$.
\end{defi}

The set of candidates for a reference class is the largest possible reference class (in sense of cardinality).
It includes all objects that could potentially be a member of the reference class and additionally serves as a market climate reference class that captures the overall market sentiment for the time period of candidate firms.
The resulting ECDF may serve as an estimate of the marginal distribution neglecting any confounding variables.
The new rank deviation procedures using multiple reference variables take the backtest against a benchmark approach of \cite{maubcal15}, a simple approach using the major and industry group of a firm, and the market climate reference class.

\subsubsection{Group Approach}

It is common practice in corporate valuation to form peer groups based on an industry classification such as the standard indsutrial classification (SIC) due to the assumption that firms in the same industry are similar in terms of value determinants \citep{bhojrajlee2002, marozzi2013}.
The group approach uses the major and industry group of the SIC (first two and three digits of SIC) in a straightforward way to construct a reference class from the set of candidates.
In both cases, all candidate firms that are in the same major or industry group, respectively, as the initial firm are members of the reference class.
Thus, membership in the same major or industry group is said to fulfill the assumption of sufficient similarity.

\subsubsection{Mauboussing and Callahan (2015)}

\cite{maubcal15} propose to base sufficient similarity on the single reference variable real sales to construct 11 subsets from the set of candidates as potential reference classes that do not depend on the specific forecast challenge.
They sort the candidate firms by real sales in 10 deciles as well as an 11th subset for the top one percentile.
For any forecast challenge we assess if the initial firm's real sales is higher than the $99\%$ percentile of the candidates' real sales.
If so, the top one percentile of candidate firms in terms of real sales constitutes the reference class.
If not, we assess the decile which includes the initial firm's real sales and this decile becomes the reference class.
The potential reference classes are constructed independently of the forecast challenge and pose the risk of missing similar firms if the initial firm's real sales is close to the boundary of one of the subsets (c.f. Figure \ref{fig:reference_scheme} and its discussion in Section \ref{subsec:theory}).

\subsubsection{Rank Deviations}
We introduce a novel method using rank deviations that assesses similarity of candidate firms based on multiple reference variables and extends the approach by \cite{theisingetal2023} which uses an arbitrary but single reference variable.
Time series data in discrete time can be incorporated by treating each point in time as an additional reference variable.
The novel method is rank-based to mitigate skewness effects as well as outlier influence on the selection and constructs custom reference classes seperately for each forecast challenge, i.e. each inital firm here.
Thus, there is no risk of neglecting similar firms as opposed to \cite{maubcal15} who a priori fix the potential reference classes.

Sufficient similarity is measured based on ranks and a size parameter $c \in (0,1)$ that controls the size of the reference class as a fraction of the candidate set and thereby exploits the continuity of reference variables.
Consequently, the parameter $c$ determines which of the candidates' reference variables $X_{j,\zeta :s}$ lie closely enough to the initial firm's reference variable $X_{i,\tau :t}$ to be a member of the reference class and assesses for which candidate firms $j$ at time $s$ the value $d (X_{i,\tau :t}, X_{j,\zeta :s})$ is considered as \textit{small}.
The case of a single reference variable illustrates the method.
We select a fraction $c$ of candidates $(j,s)\in C$ for the reference class with the least absolute rank deviation $\vert R(X_{i,t}) - R(X_{j,s})\vert $ as sufficient similar to the initial firm, where the rank function $R:\mathbbm{R} \to [1,\vert C\vert + 1]$ calculates the rank of a single reference variable in the set of candidate firms and the initial firm.

\begin{figure}
	\centering
	\includegraphics[width=\textwidth]{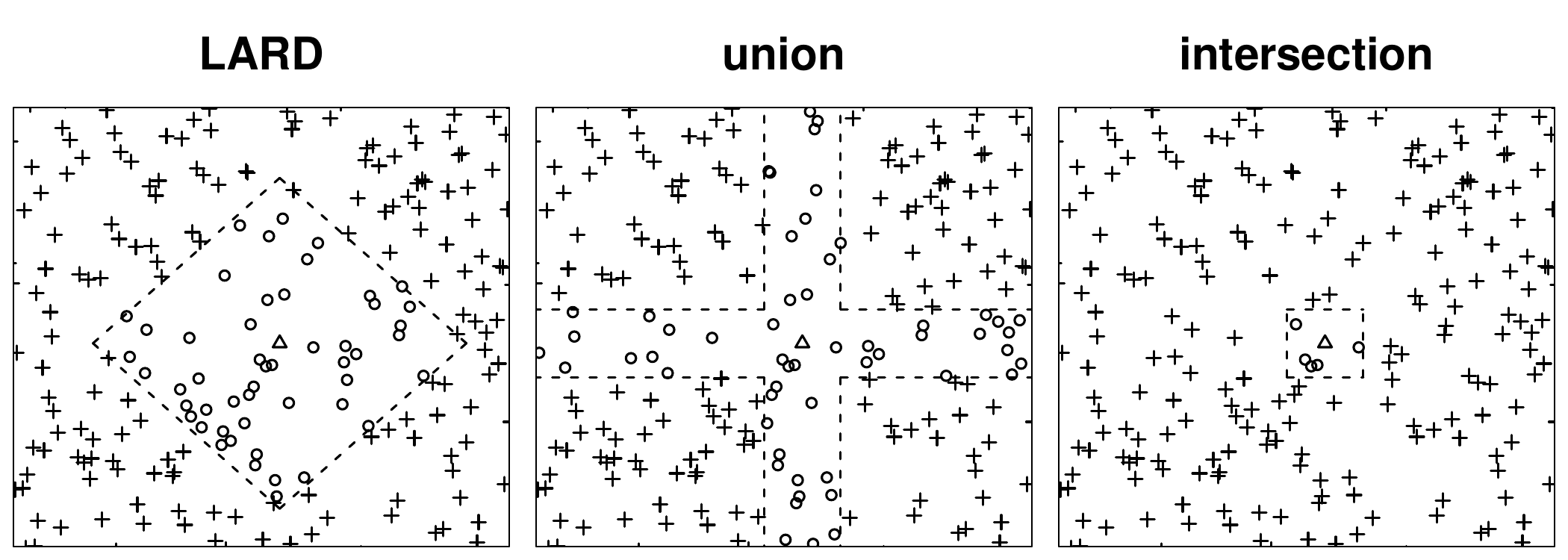}
	\caption{Illustration of the three rank deviation methods for reference class selection based on $\kappa = 2$ reference variables on them same candidate set with size $N=500$ and reference class size $c = 0.1$. Ranks of the reference variables between 100 and 400 are displayed on the horizontal and vertical axes. The triangle shows the initial firm, circles are selected as reference class members and crosses are the remaining observations.}
	\label{fig:rkdev_bivar}
\end{figure}

We propose three ways of combining $\kappa > 1$  reference variables by intersecting or unifying the reference classes obtained from several single reference variables or by using the candidate firms that have least absolute rank deviation (LARD) inspired by \cite{kkp17}.
The two set-theoretic operations both involve first constructing $\kappa$ reference classes based on each reference variable seperately.
On the one hand, we combine the reference classes by intersecting them with the possibility of having few or none observations left.
Constructing the initial reference classes with an adjusted $c_{\text{inter}} = \min \{c \kappa , 0.25 \}$ may avoid an insufficient amount of remaining observations.
On the other hand, we combine the reference classes by union where selecting too many candidates may be solved by constructing the initial reference classes with an adjusted $c_{\text{union}} = c / \kappa$.
Unifying the reference classes has the additional advantage that not all reference variables must be observed for each reference class candidate.
The application of LARD requires a ranking of candidate firms and inital firms according to each reference variable resulting in rank vectors $r_{i,t} = \mathcal{R}(X_{i,\tau :t})$ for the initial case and $r_{j,s} = \mathcal{R}(X_{j, \zeta :s})$ for all candidates $(j,s)\in C$, where $\mathcal{R}$ is the rank function applied on each entry of the reference variables seperately.
Then, the fraction $c$ of observations with the least absolute rank deviation $d_{j,s} = \vert r_{j,s} - r_{i,t} \vert$ in $L_1$ norm $\vert \vert d_{j,s} \vert \vert_1$ constitutes the reference class.
The algorithm is related to the $k$ nearest neighbors algorithm where $k = cN$ is chosen relative to the number of candidates $N$ and proximity is measured by $L_1$-norm of ranks without subsequent regression or classification but with a distribution forecast.
Other norms could be used but $L_1$ is a natural choice applied on ranks.
Further, intersecting $\kappa$ reference classes is related to the supremum norm of the vectors of absolute rank deviations $d_{j,s}$ and the union of $\kappa$ reference classes is comparable to selecting reference classes by the minimum entries of $d_{j,s}$ (c.f. Figure \ref{fig:rkdev_bivar} for both).

Using rank deviation for a single reference variable is equivalent to the procedure in \cite{theisingetal2023} where candidate firms are ordered by a single variable and the fraction $c$ of candidates closest to the initial firm's observation consitute the reference class.
Let $\hat{F}_{\text{cand}}$ be the ECDF of all candidate firms and $\hat{F}^{-1}_{\text{cand}}$ be the associated empirical quantile function of all candidate firms.
Then, all candidates $(j,s) \in C$ with $\vert \hat{F}^{-1}_{\text{cand}}(X_{i,t}) - \hat{F}^{-1}_{\text{cand}}(X_{j,s})\vert \leq c/2$ are chosen for the reference class.
If the initial case's reference variable is at the tail of the candidates' distribution, \cite{theisingetal2023} choose the top or bottom fraction $c$ of the candidates regarding the reference variable if $\hat{F}^{-1}_{\text{cand}}(X_{i,t}) > 1 - c/2$ or $\hat{F}^{-1}_{\text{cand}}(X_{i,t}) < c/2$, respectively.
This is identical to selecting the fraction $c$ of candidate firms with least absolute rank deviation.

\subsubsection{Principal Component Analysis Rank Deviation}

In order to use information from a large number of co-variates, we first apply principal component analysis (PCA) to reduce the dimensionality of the problem and then use the rank deviation procedures on the rotated data to identify the reference class.
Combing several reference variables by LARD is related to the $k$ nearest neighbors algorithm which is used in algorithmic pipelines with PCA, e.g., in facial recognition \citep{marcialisroli2004,parvthur06}.
Although the union procedure allows different sets of reference class candidates for each co-variate, this is no longer the case for PCA preprocessing.
Constructing the reference class based on principal components (PCs), all variables must be available in order to rotate the original data matrix.

PCA is carried out on the original reference variables, a transformed set of reference variables $\mathcal{X}_{j, \zeta :s}$ or a subset $C'$ of the candidate set $C$ obtained by one of the following four initial transformations: 
a) no inital transformation, i.e. $\mathcal{X}_{j, \zeta :s} = X_{j, \zeta :s}$ and $\mathcal{X}_{i, \zeta :t} = X_{i, \tau :t}$; 
b) use the initial transformation $\mathcal{X}_{j, \zeta :s} = X_{j, \zeta :s}^{1/5}$ and $\mathcal{X}_{i, \tau :t} = X_{i, \zeta :t}^{1/5}$ to mitigate skewness effects in the data;
c) compute ranks $\mathcal{X}_{j, \zeta :s} = \mathcal{R}(X_{j, \zeta :s})$ and $\mathcal{X}_{i, \tau :t} = \mathcal{R}(X_{i, \tau :t})$ for each reference variable seperately;
or d) trim the data for each reference variable across the candidate set by $2.5\%$ on each tail and then reduce the candidate set $C$ by all candidates that get trimmed in at least one reference variable such that all observations are complete in the subset of remaining candidates $C'\subset C, \vert C' \vert = N'$.

Let $\mathcal{X}$ be the $N\times \kappa$ (or $N'\times \kappa$) data matrix containing the potentially transformed reference variables from the set of (remaining) candidates $C$ (or $C'$, respectively), and let $W$ be the $\kappa \times \kappa$ weight matrix whose columns are the eigenvectors of the correlation matrix $(\mathcal{X}'\mathcal{X})^{-1/2}\mathcal{X}'\mathcal{X}(\mathcal{X}'\mathcal{X})^{-1/2}$.
The transformation $\mathcal{X} W$ maps the $\kappa$ reference variables $\mathcal{X}$ to a new $\kappa$-dimensional space.
As we use the correlation matrix, the largest variance by scalar projection of $\mathcal{X}$, standardized to variance 1 for each column, lies on the first column of $\mathcal{X}W$, the second largest variance lies on the second column of $\mathcal{X}W$ and so forth up to the smallest variance on the last column \citep[p. 30]{jolliffe2002}.
Finally, we use $W_{1:L}$, the matrix of the first $L$ columns of $W$, to perform the rank deviation based reference class selection on the dimension reduced matrix obtained as $\mathcal{X}_L = \mathcal{X} W_{1:L}$ for initial transformations a) - c), and $\mathcal{X}_L = X W_{1:L}$ for initial transformation d), where $X$ is the $N \times \kappa$ matrix of untransformed reference variables.
Naturally, we need to calculate $\mathcal{X}_{i,t} W_{1:L}$ for the (potentially) transformed reference variables of the inital firm to assess the rank deviations.

The number of principal components $L$ is chosen by different strategies.
On the one hand, for the sake of interpretation we investigate simply using two or three PCs.
On the other hand, data driven criteria select the number of PCs that explain at least 75\% or 90\% of the total variability, or that explain more variability than the mean variability across all PCs $\mathrm{Var}_{\mu}$.

\section{Performance of Distributional or Probabilistic Forecasts}

The resulting distributional information serve as forecasts and the suitability of reference classes is assessed by the distributional forecast accuracy.
Typically, forecast performance is evaluated by measuring the distance between a forecast and the realized outcome according to a loss function, taking the average loss across all forecast instances and comparing forecast models by their mean loss.
Distributional forecasting renders this method infeasible as the realized outcome is not a cumulative distribution function and a distance to the forecast cannot be calculated.
We need to evaluate the forecast performance with measures for this specific setting.
In line with the prequential principle \citep{dawid1984} we base the evaluation of the forecast model only on forecasts it actual performed and the subsequent realized outcomes in a backtest on historical data.

Here, we must evaluate the forecast quality based on the forecast distribution $F^*_{i,t;h}$ and the observed outcome $y_{i,t+h}$ of $Y_{i,t+h}$ with distribution function $F_{i,t;h}$ for a fixed forecast horizon $h$.
\cite{dawid1984} and \cite{dieboldetal1998} propose to use the transformation 
\begin{align}\label{PIT}
	F^* (y_{i,t+h}) := n^{-1} \sum_{(j,s)\in R} \mathbbm{1}\{ Y_{j,s+h} \leq y_{i,t+h}\} \approx \mathbbm{P} (Y_{i,t+h} \leq y_{i,t+h}) =: F(y_{i,t+h})
\end{align}
for forecast evaluation, where $n=\vert R\vert$.
For an ideal forecast $F^*_{i,t;h}=F_{i,t;h}$, \eqref{PIT} holds exactly and $F^* (y_{i,t+h})$ is the probability integral transform (PIT) and thus uniformly distributed on $[0,1]$.
Assuming a good forecast, \eqref{PIT} should at least hold approximately which makes the approximate uniformity of $F^* (y_{i,t+h})$ a necessary condition for a good forecast.

Repeatedly obtaining $F^* (y_{i,t+h})$ in a backtest for multiple individuals $i$ and points in time $t$ results in a sample of PIT values $\{p_k\}_{k=1,\dots, m}$ in the interval $[0,1]$, where $m$ is the number of forecast instances during the backtest. 
If the approximation of the distribution is valid, we approximately have realizations from a uniform distribution on $[0,1]$. 
The PIT is useful for absolute assessment whether a predictive distribuion is suitable by diagnosing misspecification \citep{dieboldetal1998, gneitingetal2007, heldetal2010} because uniformity of the PIT values is essentially calibration \citep[probabilistic calibration in][]{gneitingetal2007} and refers to the statistical consistency between observations and the respective distributional forecast.
To assess the forecast ability of the different algorithms and reference variables, we consider measures that determine how close this approximation is by checking if it is reasonable to maintain the hypothesis that $\{p_k\}_{k=1,\dots, m}$ stem from a uniform distribution.

For a finite number of quantile levels $\{\alpha_j\}_{j=1,\dots ,l}$ we rank models by a quantile comparision through the absolute difference between the quantiles of $\{p_k\}_{k=1,\dots, m}$ and the quantiles of the uniform distribution on $[0,1]$.
These differences are summed up and ranked.
The absolute quantile difference $\Delta_{\text{q}}$ as used in \citet{theisingetal2023} is bounded and enables us to easily calculate an interpretable mean deviation from the theoretical quantiles but is not adjusted for sample size.
Admittedly, independent of the number of quantiles there are plenty of distributions that have the same quantiles as the uniform distribution.
However, we are more interested in giving suitable reference classes and if a practitioner is particularly interested in certain quantiles of the distribution and not so much in anything else, the absolute quantile difference is feasible.
The free choice of quantile levels enables a flexible approach to highlight certain areas of the distribution that researchers or forecasters are interested in the most.
A visual inspection of histograms of the PIT values is a common forecast assessment \citep{hamill2001} and equivalent to the absolute quantile difference with bins chosen according to the quantiles. 
But while the PIT histogram might be handy if a forecaster only considers a handful of models and or variable sets the visual inspection remains qualitative in nature and is infeasible for large scale model and variable selections.

Statistical goodness-of-fit tests for uniformity are not applicable in this particular backtest as sample sizes vary between $100,000$ and $300,000$, depending on hyper parameters, and most p-values would be very small or even get reported as $0$ by software \citep[see][]{theisingetal2023}.
In this case, even the smallest forecast errors cause what \cite{berkson38} pointed out: ``Any consistent test will detect any arbitrary small change in the [distribution] if the sample size is sufficiently large''.
Avoiding this problem, we report the value of selected test statistics in addition to the absolute quantile difference.
The counterpart to a quantile comparision would be a $\chi^2$ goodness-of-fit test.
But to offer a different perspective, we focus on the Kolmogorov-Smirnov (KS) and Cramer-von-Mises (CvM) tests - common choices for testing the equality of the complete distribution .
Let $G_m$ be the empirical distribution function of $\{p_k\}_{k=1,\dots, m}$ and let $G_0$ be the distribution function of the uniform distribution on $[0,1]$. 
The corresponding test statistics are given by $\sqrt{m} \sup_{x\in [0,1]} \vert G_m (x) - G_0 (x)\vert$ (KS) and $m\int_{0}^{1} [G_m(x)-G_0(x)]^2\mathrm{d}F_0(x)$ (CvM).
These tests for continuous distributions might not be suitable if, by construction, the distribution forecast is based on the same number of observations for each forecast instance such that $\{p_k\}_{k=1,\dots, m}$ has a discrete distribution. 
In such cases and if $Y_{i,t+h}$ is discrete, a $\chi^2$-test is more suitable to assess performance.

Given the difficulty of forecasting corporate sales growth we are mainly interested in finding calibrated distributional forecasts if any exist, e.g. to correct potentially biased point forecasts, and do not focus on maximizing the sharpness subject to calibration by sharpness measures or proper scoring rules.
Sharpness refers to the concentration of the forecast characterized by scale parameters, is a property of the distributional forecast itself and can be measured by the width of certain confidence intervals, boxplots, quantiles of the distribution or scale parameters.
A scoring rule is a real-valued function that assigns a loss to a probabilistic forecast $F^*$ if the value $y_{i,t+h}$ is observed.
If a scoring rule is proper, the true distribution has a smaller loss than any other forecast distribution such that we check the equality of forecast distributions to the true distribution \citep{dieboldetal1998}.
Scoring rules are suitable for comparative assessment of multiple forecasting schemes if they refer to exactly the same set of forecast situations \citep{gneitingrafterty2007} which given the vast number of different reference variables and missing values in our data set is infeasible.
Providing the same set of forecast instances for each set of reference variables would distort the data set systematically and lead to potentially biased conclusions in the backtest.

\section{Data Set}

\begin{figure}
	\center
	\includegraphics[height=0.33\textheight]{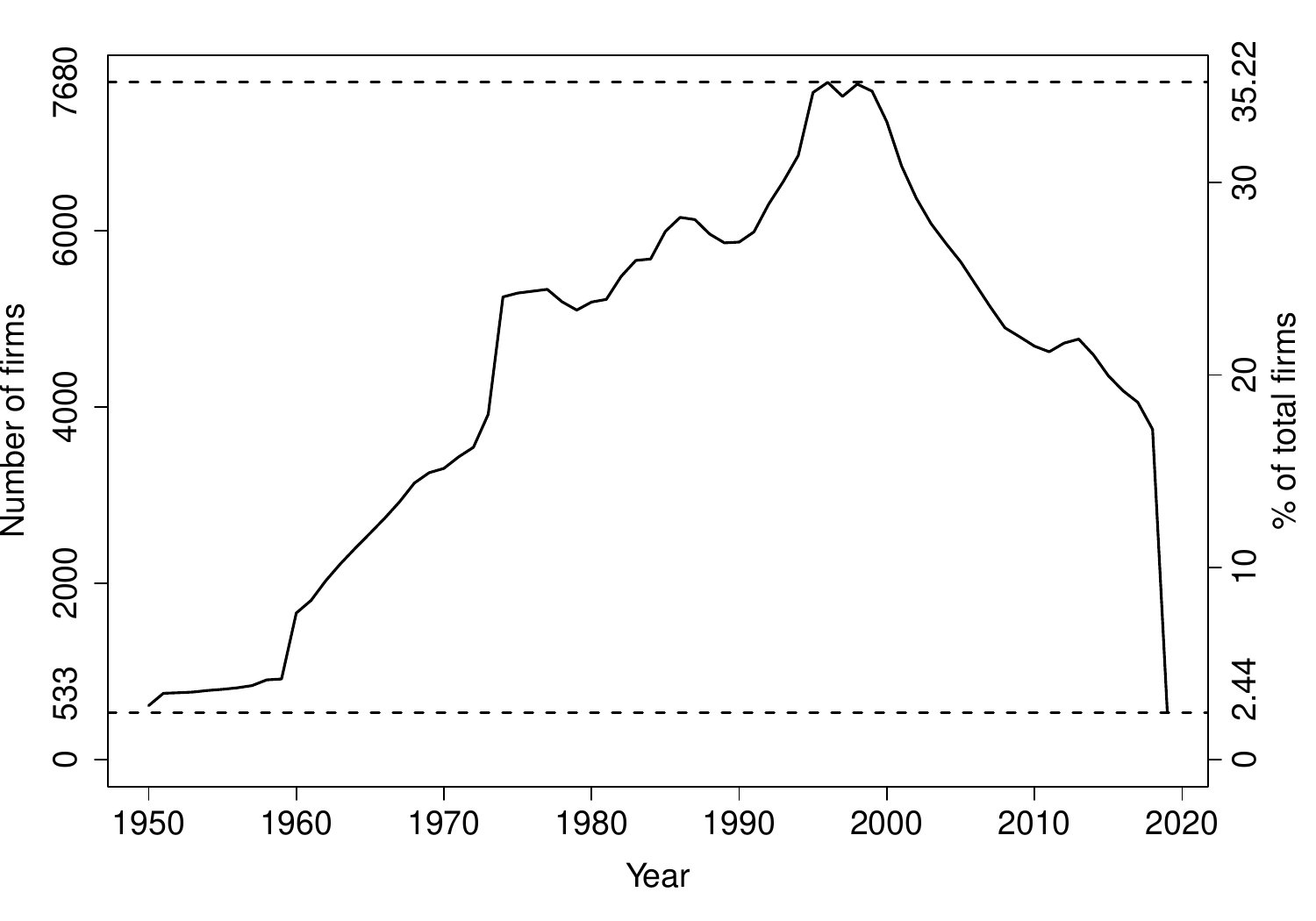}
	\caption{Number of companies over time as in \cite{theisingetal2023}. The left vertical axis shows the number of firms, i.e. observations, per year and the right vertical axis covers the number of firms as a proportion of the total number of firms.}
	\label{fig:obsperyear}
\end{figure}

In order to identify optimal combinations of reference variables and algorithm options for reference class selection within a backtest we consider the historic data set used in \cite{theisingetal2023}.
This data set consists of Compustat North America fundamentals annual data from 1950 to 2019 by \citet{compustat} limited to US firms and excluding companies from the financial and real-estate sector.
Firms with none or only one sales observation are omitted given the challenge to predict distributions of sales growth.
These data are merged with stock-exchange information from the \citet[CRSP,][]{crsp} daily stock of the University of Chicago Booth School of Business.
Variables measured in US dollar are adjusted to 1982~--~1984 US dollar using monthly inflation rate data from the consumer price index for all urban consumers (all items in US city average) by the \citet{cpi}.

\begin{figure}
	\center
	\includegraphics[height=0.33\textheight]{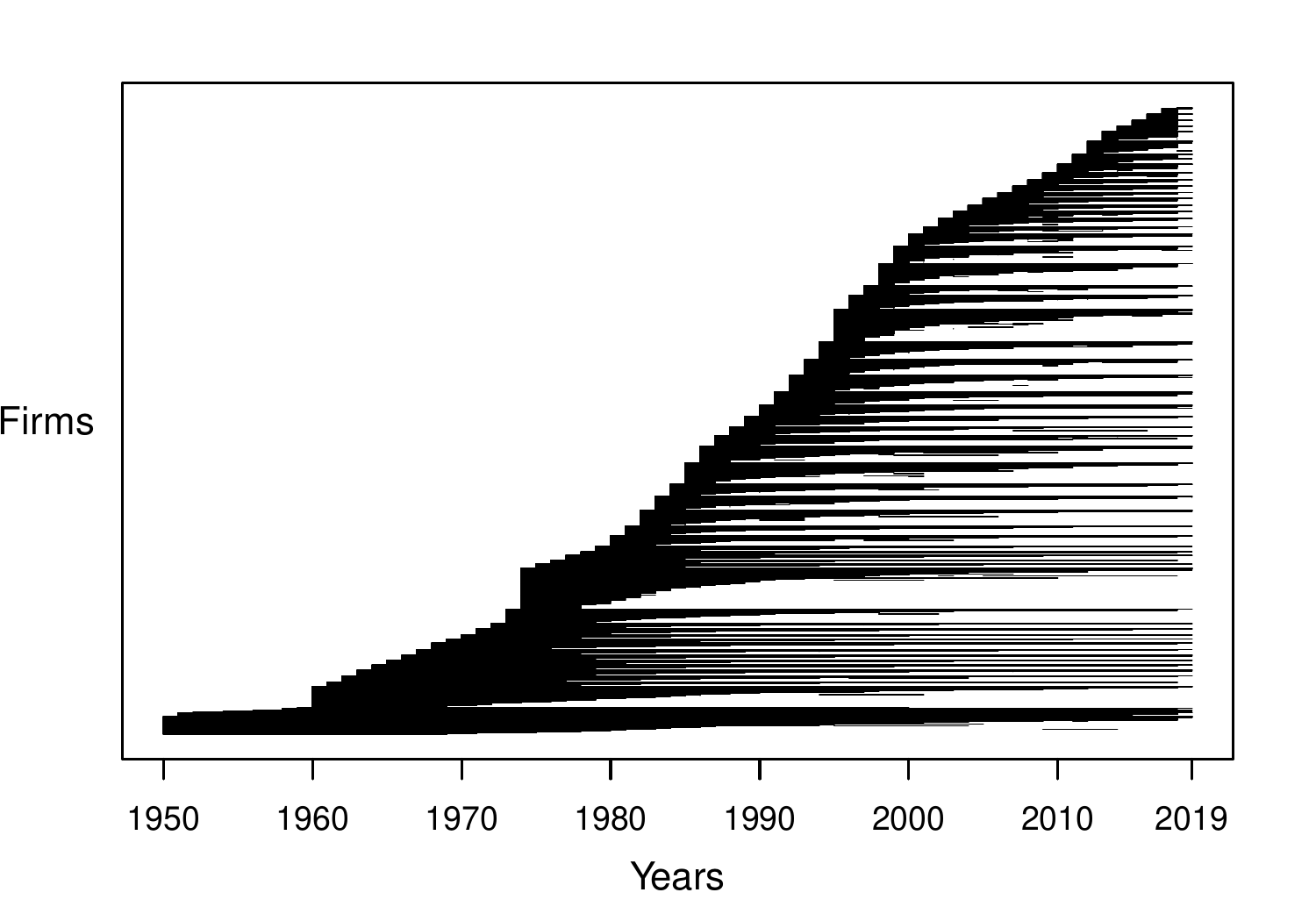}
	\caption{Time series properties of the firms as in \cite{theisingetal2023}. Each horizontal line represents one of the 21,808 firms ordered from bottom to top by three criteria: 1. the first year of appearance in the data set, 2. the number of observations of the firm, 3. the number of consecutive observations of the firm.}
	\label{fig:tslength}
	\includegraphics[height=0.33\textheight]{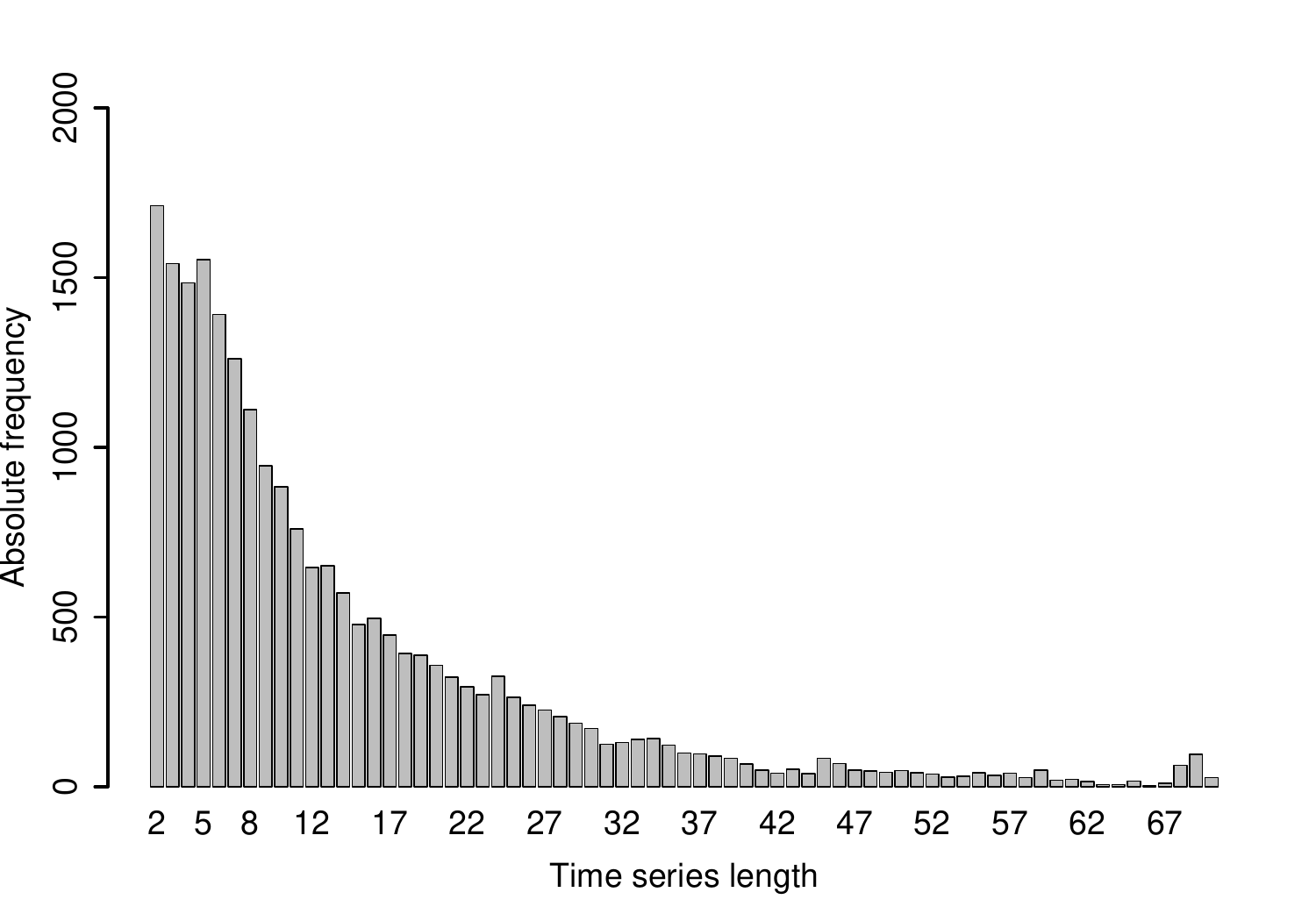}
	\caption{This barplot shows the number of observations per firm in the data set, that is the empirical distribution of time series length, as in \cite{theisingetal2023}.}
	\label{fig:tslengthbarplot}
\end{figure}

The data set contains 303,628 observations on 21,808 firms of which a total 206,221 observations on 17,099 firms provide CRSP stock exchange market information.
There is an influence of survivorship in the data set as the number of available observations per year varies considerably (c.f. Figure \ref{fig:obsperyear}) as well as the time series lengths of firms (c.f. Figures \ref{fig:tslength} and \ref{fig:tslengthbarplot}).
The survivorship rates for forecast horizons considered in our backtest are 97.25\% for one year, 89.61\% for three years, 76.12\% for five years and 48.20\% for 10 years.

\begin{table}
    \small
    \centering
    \caption{\small Description of reference variables. EBIT is the earning before interest and taxes, market cap. is the market capitalization and pp is percentage points. A variable summary can be found in Table \ref{tab:predsummary}.}
    \label{tab:predictors}
    \begin{tabular}{lll}
        Abbreviation & Reference Variable & Description \\
        \midrule
        at & total assets & in million USD \\
        opmar & operating margin & EBIT divided by sales (in \%) \\
        -- & sales & in million USD \\
        seq & shareholder equity & total assets minus total liabilities \\
         &  & (in million USD) \\
        -- & major group & first two digits of SIC, 63 groups \\
        -- & industry group & first three digits of SIC, 250 groups \\
        -- & $\beta$ & slope of regressing daily return on market \\
         &  & return \\
        P/B & price-to-book ratio & market cap. divided by shareholder equity \\
        P/E & price-to-earnings ratio & market cap. divided by net income \\
        $\text{salesGR}_{\tau}$ & $\tau$-year past sales growth & current sales divided by sales $\tau$ years ago (in \%)\\
        $\text{opmar}\Delta_{\tau}$ & $\tau$-year past operating & current operating margin minus operating \\
         & margin delta & margin $\tau$ years ago (in pp)\\
    \end{tabular}
\end{table}

\begin{table}
    \small
    \centering
    \caption{\small Summary of reference variables (ref.var) from Table \ref{tab:predictors} as in \cite{theisingetal2023}, where qu. is quantile. $\text{salesGR}$ is compound annual growth rate in \% in this table and $\text{opmar}\Delta$ is the mean annual past operating margin delta to simplify comparision across lags. The summary on major and industry groups covers the group sizes.}
    \label{tab:predsummary}
    \begin{tabular}{l|rrrrrr|r}
         Ref.Var. & 2.5\% qu. & 25\% qu. & Median & Mean & 75\% qu. & 97.5\% qu. & Missings \\
        \midrule
        at & 0.27 & 11.82 & 62.31 & 877.65 & 337.77 & 6767.24 & 2714 \\
        opmar & -827.80 & -1.19 & 6.01 & -402.68 & 12.27 & 34.49 & 18532 \\
        sales & 0.00 & 10.67 & 67.60 & 721.10 & 337.74 & 5345.51 & 0 \\
        seq & -9.65 & 3.58 & 24.00 & 319.76 & 128.97 & 2478.79 & 19811 \\
        major group & 10 & 895 & 2646 & 4819.49 & 5295 & 25617 & 0 \\
        industry group & 38 & 283 & 622 & 1214.51 & 1248 & 6793 & 0 \\
        $\beta$ & -0.28 & 0.37 & 0.77 & 0.83 & 1.21 & 2.31 & 97469 \\
        P/B & -6.00 & 0.59 & 1.34 & 2.65 & 2.57 & 11.70 & 100318 \\
        P/E & -70.39 & -3.45 & 8.34 & 11.24 & 17.69 & 104.99 & 98786 \\
        $\text{salesGR}_{1}$ & -100 & -5.39 & 4.93 & 115.70 & 19.24 & 1465000 & 31591 \\
        $\text{salesGR}_{2}$ & -100 & -4.18 & 4.55 & 17.07 & 16.33 & 19090 & 52164 \\
        $\text{salesGR}_{3}$ & -100 & -3.31 & 4.32 & 10.41 & 14.51 & 3862 & 71103 \\
        $\text{salesGR}_{4}$ & -100 & -2.71 & 4.21 & 7.90 & 13.17 & 1794 & 88572 \\
        $\text{salesGR}_{5}$ & -100 & -2.22 & 4.13 & 6.52 & 12.23 & 1019 & 104702 \\
        $\text{salesGR}_{6}$ & -100 & -1.87 & 4.05 & 5.62 & 11.44 & 609.50 & 119372 \\
        $\text{salesGR}_{7}$ & -100 & -1.55 & 4.00 & 5.02 & 10.82 & 435.80 & 132772 \\
        $\text{salesGR}_{8}$ & -100 & -1.29 & 3.98 & 4.59 & 10.38 & 333.90 & 145044 \\
        $\text{salesGR}_{9}$ & -100 & -1.06 & 3.95 & 4.28 & 9.97 & 277.10 & 156300 \\
        $\text{salesGR}_{10}$ & -100 & -0.87 & 3.91 & 4.03 & 9.58 & 205.30 & 166682 \\
        $\text{opmar}\Delta_{1}$ & -2824000 & -2.73 & 0.04 & -10.15 & 2.57 & 2823000 & 41527 \\
        $\text{opmar}\Delta_{2}$  & -1412000 & -1.96 & -0.03 & -11.85 & 1.71 & 681300 & 62660 \\
        $\text{opmar}\Delta_{3}$ & -374800 & -1.54 & -0.07 & 4.04 & 1.26 & 951200 & 81829 \\
        $\text{opmar}\Delta_{4}$ & -326200 & -1.27 & -0.08 & 3.89 & 1.00 & 691100 & 99288 \\
        $\text{opmar}\Delta_{5}$ & -260800 & -1.09 & -0.08 & 3.19 & 0.82 & 523200 & 115291 \\
        $\text{opmar}\Delta_{6}$ & -217300 & -0.95 & -0.09 & 0.42 & 0.69 & 204400 & 129585 \\
        $\text{opmar}\Delta_{7}$ & -107800 & -0.84 & -0.09 & 3.81 & 0.60 & 185700 & 142583 \\
        $\text{opmar}\Delta_{8}$ & -89290 & -0.76 & -0.08 & 2.25 & 0.53 & 190800 & 154449 \\
        $\text{opmar}\Delta_{9}$ & -81610 & -0.69 & -0.08 & 3.21 & 0.46 & 335300 & 165288 \\
        $\text{opmar}\Delta_{10}$ & -75350 & -0.64 & -0.08 & 3.44 & 0.41 & 301700 & 175265 \\
    \end{tabular}
\end{table}

As potential reference variables, we select key figures in fundamental analysis such as sales, operating margin, total assets, shareholder equity, the SIC (standard industrial classification), $\beta$, the price-to-earnings ratio and the price-to-book ratio.
We additionally construct one to 10 year past sales growth and one to 10 year past operating margin delta where the necessary past data are available.
For the group approach, we derive a firm's major and industry group from the SIC.
Table \ref{tab:predictors} describes all potential reference variables and Table \ref{tab:predsummary} summarizes them including certain quantiles, their means and the number of missing values in the data set.
Note that we display compound annual sales growth rates and annual means of operating margin deltas for comparision purposes across lags but later use non averaged sales growth rates and operating margin deltas for reference class selection.
Given the overall frequency of missing values, e.g. more than 50\% of the data for 10-year sales growth, handling only identical forecast challenges for all different combinations of reference variables would reduce the number of forecast cases too drastically and thereby generate issues regarding survivorship influence.
Thus, scoring rules are infeasible under these circumstances.
In addition to Table \ref{tab:predsummary}, negative skew occurs for reference variables operating margin, $\beta$, one and two year operating margin delta and price-to-earnings ratio, as the smallest with roughly~$-167$.
All other reference variables have a positive skew with up to roughly $400$ in the case of price-to-book ratio.
This supports the use of rank based methods to reduce skewness effects.

\begin{table}
    \center
    \caption{Compound annual sales growth rates for the whole data set as in \cite{theisingetal2023}. Mean and standard deviation are 2.5\% trimmed on both tails, the respective quantiles are in the table.}
    \label{tab:cagr-fulluniverse}
    \begin{tabular}{l|rrrr}
        Full Universe & \multicolumn{4}{c}{Base Rates}\\
        \midrule
        CAGR (\%) & 1-Yr & 3-Yr & 5-Yr & 10-Yr \\
        \midrule
        $\leq -25$ & 8.70 & 5.44 & 4.00 & 2.38 \\
        $]-25,-20]$ & 2.19 & 1.69 & 1.28 & 0.68 \\
        $]-20,-15]$ & 3.18 & 2.65 & 2.13 & 1.37 \\
        $]-15,-10]$ & 4.53 & 4.27 & 3.71 & 2.68 \\
        $]-10,-5]$ & 7.06 & 7.28 & 7.11 & 6.12 \\
        $]-5,0]$ & 10.92 & 13.20 & 14.29 & 15.64 \\
        $]0,5]$ & 13.59 & 17.82 & 21.17 & 27.25 \\
        $]5,10]$ & 11.65 & 14.33 & 16.34 & 20.09 \\
        $]10,15]$ & 8.24 & 9.06 & 9.70 & 9.95 \\
        $]15,20]$ & 5.65 & 5.86 & 5.77 & 5.38 \\
        $]20,25]$ & 4.08 & 3.95 & 3.61 & 2.92 \\
        $]25,30]$ & 3.05 & 2.71 & 2.54 & 1.76 \\
        $]30,35]$ & 2.31 & 2.04 & 1.73 & 1.14 \\
        $]35,40]$ & 1.78 & 1.54 & 1.26 & 0.69 \\
        $]40,45]$ & 1.46 & 1.17 & 0.93 & 0.48 \\
        $> 45$ & 11.58 & 6.99 & 4.42 & 1.46 \\
        \midrule
        mean & 10.62 & 7.01 & 5.75 & 4.62 \\
        \midrule
        median & 4.93 & 4.32 & 4.13 & 3.91 \\
        \midrule
        std & 32.30 & 19.08 & 14.21 & 9.20 \\
        \midrule
				$q_{0.025}$ & -60.01 & -44.75 & -36.52 & -23.91 \\
        \midrule
				$q_{0.975}$ & 206.31 & 95.19 & 62.75 & 35.85 \\
        \midrule
    \end{tabular}
\end{table}

\begin{figure}
	\center
	\includegraphics[height=0.33\textheight]{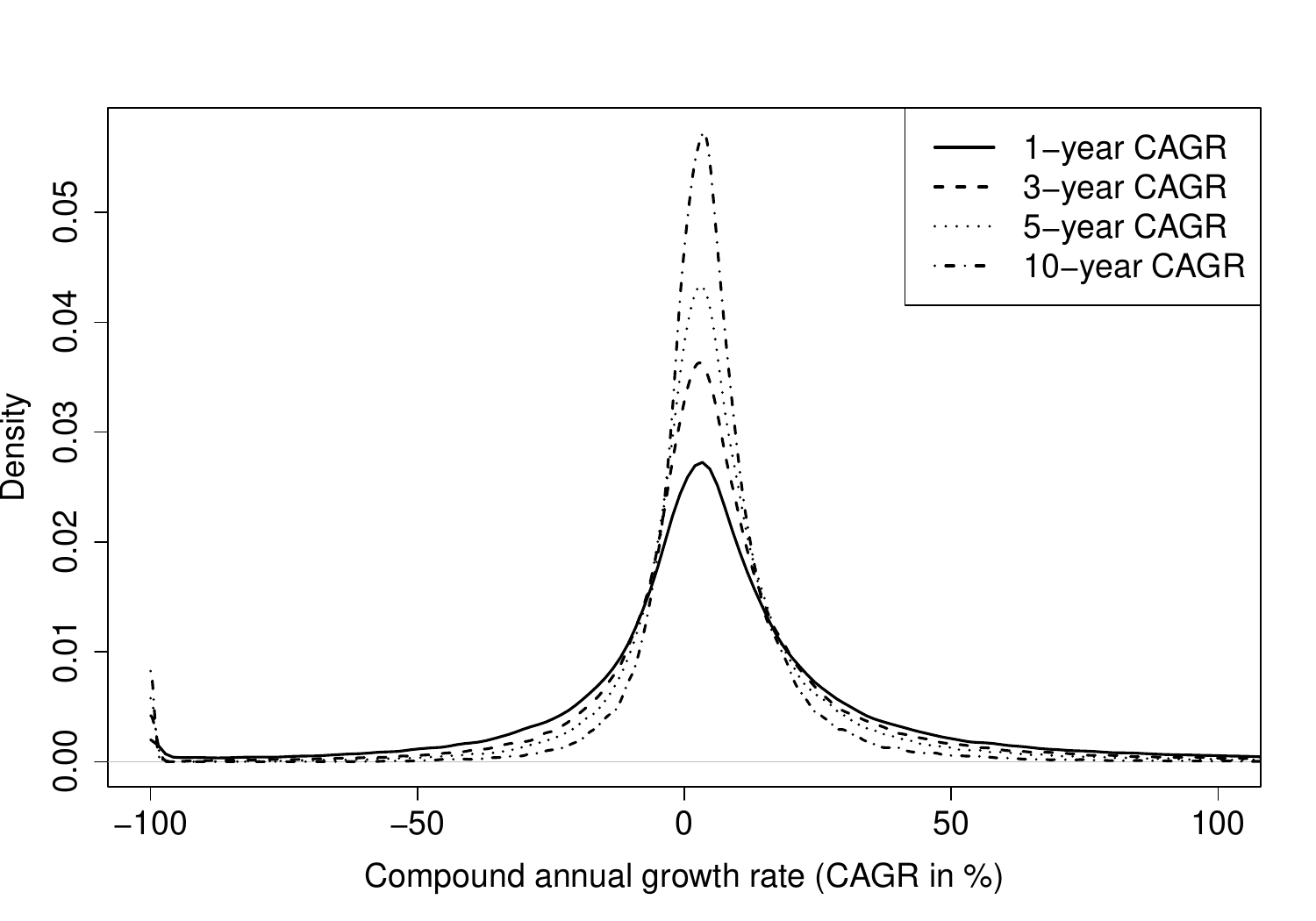}
	\caption{Estimated densities of compound annual sales growth rate for horizons one, three, five and 10 years as in \cite{theisingetal2023} using Gaussian kernel with Silverman's rule of thumb as bandwidth on support $[-100,\infty )$.}
	\label{fig:cagr-dens}
\end{figure}

In the backtest, we consider one-, three-, five- and 10-year forecast horizons, construct reference classes for sales growth and assess their quality by the forecast distributions.
Table \ref{tab:cagr-fulluniverse} shows the estimated density of historical sales compound annual growth rate (CAGR) for the complete data set.
Here, the tails of the distribution get lighter, the (2.5\%-trimmed) standard deviation decreases, the (2.5\%-trimmed) mean shifts towards the median and the distribution gets more centered the longer the forecast horizon is, as it is visible in Figure \ref{fig:cagr-dens} as well.
By a 2.5\%-trimmed mean or standard deviation we are referring to the arithmetic mean or standard deviation, respectively, where the largest 2.5\% and the smallest 2.5\% of the data are excluded.\footnote{For a vector of sorted observations $\{x_i\}_{i=1,\dots, n}$ we compute any $\alpha$-trimmed measure, $0< \alpha <1$, based on the trimmed vector of observations $\{x_i\}_{i=[\alpha n]+1,\dots , n - [\alpha n]}$, where $[\cdot ]$ is the floor function.}
The (2.5\%-trimmed) means of sales CAGR are larger than the respective medians since the growth rates are left bounded, right unbounded and we observe a considerable amount of high values that additionally render untrimmed mean and standard deviation little informative.
Using trimmed versions of these measures curbs the influence of these outliers and keeps them informative.
Summary statistics of sales CAGR can be found in Table \ref{tab:predsummary} as future and past growth rates in the full data set have identical distributions.

\section{Backtest}\label{sec:backtest}

\begin{table}
	\center
	\caption{General hyper parameters for reference class selection.}
	\label{tab:parameters}
	\begin{tabular}{l|l|l}
		Name & Abbreviation & Description \\
		\midrule
		reference variables & ref.var. & see Tables \ref{tab:predictors} \\
		class size & $c$ & relative size $\in \{0.050, 0.025, 0.010\}$ \\
		window length & $w$ & number of past years $\in \{5, 10, 20, 30\}$
	\end{tabular}
\end{table}

By means of a backtest we evaluate the performance of the new rank based methods to construct reference classes of sales growth rate forecasts on a data set ranging from $1950$ to $2019$ for forecast horizons 1, 3, 5 and 10.
Apart from the novel rank deviations and PCA rank deviations based reference class selection, we include the market climate reference class, the group approach and the approach by \cite{maubcal15} as benchmarks.
Algorithm parameters are shown in Tables \ref{tab:parameters} and \ref{tab:algorithms} and, for fixed reference variables, rank deviation has 60 possible option combinations and there are 1,200 possible combinations for PCA rank deviation.
Backtesting as a special case of cross-validation in time series settings is out-of-sample by construction and involves, in our case, distributional forecasts based on reference classes for inital firms on a historic data set. 

\begin{table}
	\center
	\caption{Different algorithm options and parameters (see Table \ref{tab:parameters}). Param. is parameters, \#{}PC is the criterioin to choose principal components and MC is \cite{maubcal15}. Combination methods union and intersection additionally offer to correct the reference class size by the number of reference variables. Transformation for PCA rank deviation is the pre PCA transformation, the subsequent transformation uses ranks.}
	\label{tab:algorithms}
	\begin{tabular}{l|l|l|l|l|l}
		Algorithm & Param. & Ref.Var. & Transformation & Combination & \#{}PC \\
		\midrule
		market climate & $w$ & - & - & - & - \\
		group approach & $w$ & SIC & first two or & - & -\\
		& & & three digits & & \\
		MC & $w$ & sales & - & - & - \\
		rank deviations & $w, c$ & all & ranks & LARD, union, & - \\
		& & & & intersection & \\
		PCA rank & $w, c$ & all & none, ranks, & LARD, union, & 2, 3, 75\%\\
		deviations & & & trim, $x^{1/5}$ & intersection & 90\%, $\mathrm{Var}_{\mu}$
	\end{tabular}
\end{table}

\begin{figure}
	\centering
	\includegraphics[width=.95\textwidth]{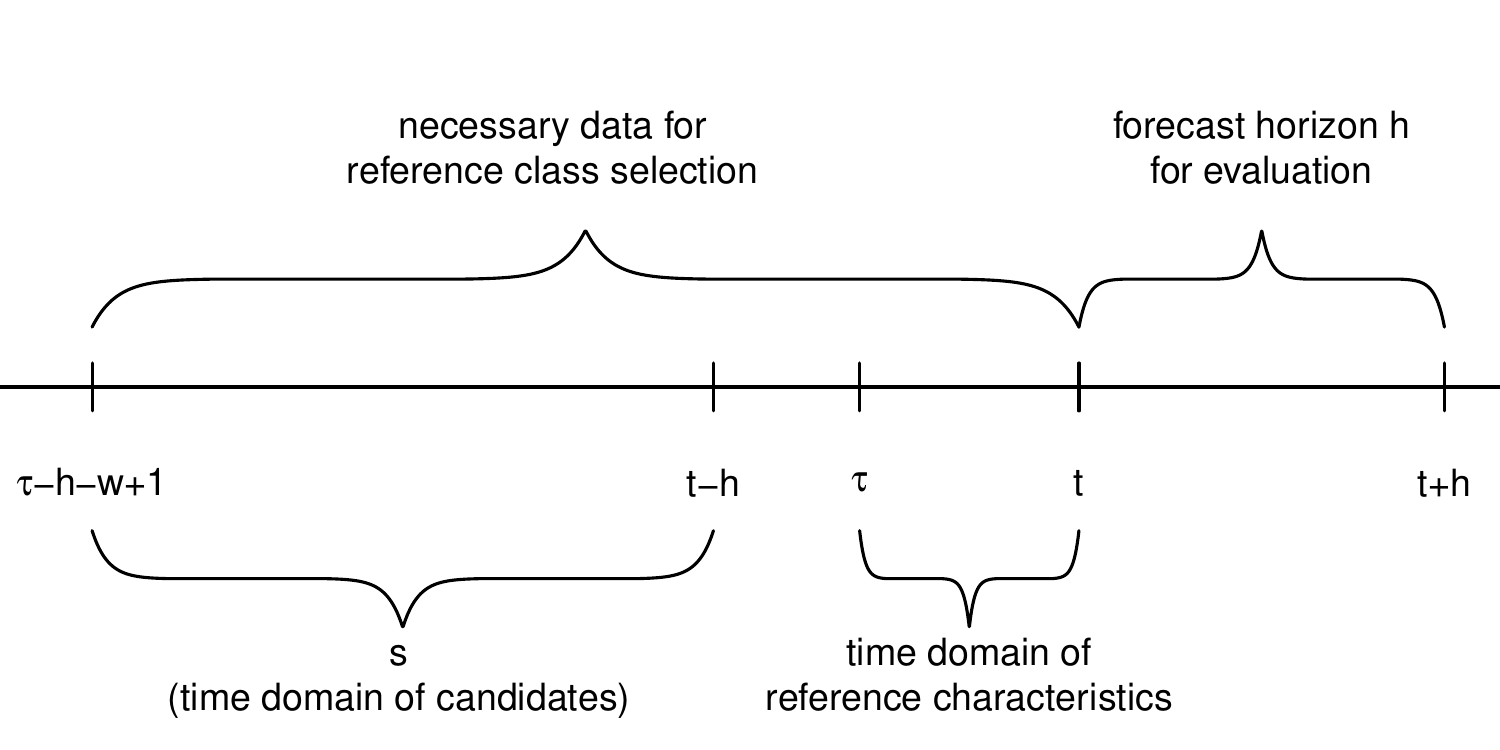}
	\caption{Illustration of the backtest timeline. Note, $\tau \leq t-h$ is possible as well.}
	\label{fig:backtest_scheme}
\end{figure}

Observations from the data set qualify for the backtest as initial firms depending on the forecast horizon $h$ and the window length $w$ that controls the number of past years to provide candidates for the reference class.
Assuming that at time $t$ all information of the financial year $t$ is available, each firm $i$ at each available point in time $t$ is an initial case if all used reference variables are observed, and if the full timeframe of candidates as well as the $h$-year future sales growth is available, i.e. $1950 + w + h - 1 \leq t \leq 2019 - h$ and firm $i$ is in the data set at time $t + h$.
For a fixed $t$, all firms $j$ at times $s$ serve as candidates for the initial case's reference class if they are within the window period of candidates, i.e. $t-h-w+1 \leq s \leq t-h$, and if the reference variables and $h$-year sales growth are available (see Figure \ref{fig:backtest_scheme}).
Thus, the data set is restricted to all observations without missing values with respect to sales growth rate and the used reference variables with an exception for the rank deviation procedure and union of single reference classes.
Depending on the set of candidates, the size parameter $c$ and the algorithm we construct a reference class of at least 20 elements.

For each initial firm $(i,t)$ we obtain a reference class, derive the base rates as the ECDF of sales growth rates of the reference class elements $\{y_{j,s+h}\}_{(j,s)\in R}$ and evaluate the ECDF at the realized sales growth rate $y_{i,t+h}$ of the inital firm.
Thus, we obtain a forecasted probability of being less or equal to the inital firm's realized sales growth.
In total, all inital cases produce a sample of PIT values $\{p_k\}_{k=1,\dots, m}$ and we use them to assess the forecast method and choice of reference variables.
The sample size $m$ depends on the forecast horizon, the window length, the algorithm and the availability of reference variables.
%If the distributional forecast of sales growth is calibrated, we roughly have realizations from a unifrom distribution on $[0,1]$.
As a measure of accuracy, we calculate the differences of the 1\%, 5\%, 10\%, 25\%, 50\%, 75\%, 90\%, 95\% and 99\% quantiles of $\{p_k\}_{k=1,\dots, m}$ and of the uniform distribution on $[0,1]$, respectively, and sum up the absolute quantile difference $\Delta_{\text{q}}$ with $0 \leq \Delta_{\text{q}} \leq 4.5$ for our choice of quantile levels.
The choice of quantiles is motivated by an emphasis on the distribution tails in contrast to a set of equidistant quantile levels.
Further, we report KS and CvM test statistics as accuracy measures of the whole distributional approximation.

\subsection{Variable and Model Selection Procedure}\label{subsec:varselection}

Finding appropriate reference classes is in essence a variable and model selection problem.  
We systematically explore which reference variables contain information for a calibrated distributional forecast based on rank deviations by a forward selection and brute force approaches on all contemporaneous reference variables and on selected reference variable subsets.
PCA rank deviation applied on selected reference variable subsets completes the procedure.
Forecast horizons investigated are 1, 3, 5, and 10 years and for each horizon we backtest $67,420$ different variable and model combinations.

For a systematic backtest of the rank deviation algorithm on multiple reference variables we use a forward selection to reduce the number of possible reference variable combinations.
We begin with the best three reference variables according to $\Delta_{\text{q}}$ from results in \cite{theisingetal2023} for each forecast horizon and combine them with each of the remaining reference variables using 60 different algorithm options.
We continue with the three best reference variable pairs from the previous stage with two reference variables and combine them with each of the remaining reference variables and all possible algorithm options.
Then, we repeat this for every stage by choosing the three best sets of reference variables from the previous stage and combine them with another reference variables for all possible options. 
We stop if adding another reference variable does not improve the results anymore.
However, if the forward selection comes to an early halt we continue anyways in order to protect against finding a local minimum.
The forward selection terminates when results for none of the forecast horizons improve, that is after using six reference variables.
Thus, we backtest $21,780$ different combinations.

Further, we explore using exclusively contemporaneous reference variables due to the lower data requirements opposed to lagged variables that are chosen by the forward selection.
Therefore, we brute force all combinations of seven contemporaneous reference variables (except SIC) and additionally combine the contemporaneous balance sheet variables and the full set of contemporaneous variables with up to 1, 3, 5 and 10 year lagged sales growth and operating margin delta, respectively.
This results in 127 variable combinations for the brute force approach and eight sets of contemporaneous variables with different degrees of lagged variables %, each paired with 60 algorithm options 
and, thus, $8,100$ different combinations for each forecast horizon.

Investigating the benefits of dimension reduction, we use PCA on different sets of reference variables before applying the rank deviation methods.
The $31$ reference variable subsets under consideration are all four contemporaneous balance sheet variables, all contemporaneous variables, i.e. with added financial market variables, each of these contemporaneous variable sets combined with up to 1, 3, 5 and 10 year lagged variables, the combination of the 4, 5 and 6 best reference variables, respectively, from the single reference variable approach, and each at a time the three best sets of 4, 5, and 6 best reference variables from the brute force approach and from the forward selection.\footnote{Results for these variable subsets can be found in Tables \ref{tab:resh1forward} - \ref{tab:resh10brute} in Appendix \ref{app:refvartables}.}
This results in $37,200$ different combinations for each forecast horizon.

As a benchmark, the single reference variable approach of \cite{theisingetal2023} is analyzed with regard to 27 reference variables with 12 algorithm options, resulting in 324 different combinations.
While the market climate approach and the method of \cite{maubcal15} only depend on the window sizes, the group approach uses two reference variables and four different window sizes.
In total, there are 340 different benchmark combinations for each forecast horizon.

\subsection{Results of Backtest}\label{subsec:resultsbacktest}

Tables \ref{tab:resh1rd} - \ref{tab:resh10rd} show a selection of our results\footnote{Full results are available upon request.} on forecast horizons one, three, five and 10 years, each ranked by $\Delta_{\text{q}}$ to compare the novel methods to results from \cite{theisingetal2023}.
The rank deviation (RD) results reported are the three best overall combinations, the best combination of contemporaneous reference variables and for both, in view of the necessary data, the best combination using a five- and 10-year window, respectively.
For PCA rank deviation (PCARD) we show the same selection of results as for RD.
We additionally report the best market climate window, the best results for the group approach, the best window for the method in \cite{maubcal15} and the best single reference variable overall as well as for a five- and 10-year window.
Some of these cases coincide, thus, each table has at most 22 rows.
We give details on algorithm options according to Tables \ref{tab:parameters} and \ref{tab:algorithms} and on reference variables (see Table~\ref{tab:predictors}).\footnote{Here, contemp. refers to all contemporaneous reference variables and in subscript $s$:$t$ is $\{s,\dots ,t\}$.}

\begin{table}
    \center
    \caption{Comparison of reference variables (ref. var.) and algorithms for forecasting one-year ahead sales growth. Alg. is algorithm, transf. is pre PCA transformation, comb. is combination of reference variables and cor. is correction.}
    \label{tab:resh1rd}
    \footnotesize
    \begin{tabular}{llllllll|rrr}
        Alg. & Ref. Var. & Transf. & \#{}PC & Comb. & Cor. & $w$ & Size & $\Delta_{\text{q}}$ & KS & CvM \\
        \midrule
        PCA & contemp., & ranks & 3/ & union & yes & 30 & 0.01 & 0.0045 & 1.2002 & 0.1703 \\
        & $\text{salesGR}_{1}$, &  & $\mathrm{Var}_{\mu}$ &  &  &  &  &  \\
        & $\text{opmar}\Delta_{1}$ &  &  &  &  &  &  &  \\
        PCA & contemp., & ranks & 3/ & union & yes & 30 & 0.01 & 0.0046 & 1.3626 & 0.1857 \\
        & $\text{salesGR}_{1}$, &  & $\mathrm{Var}_{\mu}$ &  &  &  &  &  \\
        & $\text{opmar}\Delta_{1}$ &  &  &  &  &  &  &  \\
        PCA & contemp., & ranks & 3/ & union & yes & 30 & 0.025 & 0.0047 & 1.4200 & 0.1755 \\
        & $\text{salesGR}_{1}$, &  & $\mathrm{Var}_{\mu}$ &  &  &  &  &  \\
        & $\text{opmar}\Delta_{1}$ &  &  &  &  &  &  &  \\
        RD & $\text{salesGR}_{3, 5:7}$, & -- & -- & union & no & 30 & 0.05 & 0.0065 & 0.7191 & 0.0407 \\
        & $\text{opmar}\Delta_{5}$ &  &  &  &  &  &  &  \\
        RD & $\text{salesGR}_{5:7}$, & -- & -- & union & no & 30 & 0.01 & 0.0066 & 0.7408 & 0.0448 \\
        & $\text{opmar}\Delta_{5}$ &  &  &  &  &  &  &  \\
        RD & $\text{salesGR}_{5:8}$, & -- & -- & union & no & 30 & 0.05 & 0.0072 & 0.6475 & 0.0698 \\
        & $\text{opmar}\Delta_{5}$ &  &  &  &  &  &  &  \\
        PCA & contemp., & ranks & 2 & union & no & 10 & 0.01 & 0.0094 & 1.6183 & 0.5917 \\
        & $\text{salesGR}_{1}$, &  &  &  &  &  &  &  \\
        & $\text{opmar}\Delta_{1}$ &  &  &  &  &  &  &  \\
        PCA & contemp. & trim & 75\% & union & no & 30 & 0.025 & 0.0102 & 1.2817 & 0.1950 \\
        single & $\text{opmar}\Delta_{6}$ & -- & -- & -- & -- & 30 & 0.025 & 0.0157 & 1.8644 & 0.8265 \\
        RD & $\beta$, P/E & -- & -- & union & no & 5 & 0.025 & 0.0158 & 2.7215 & 1.8060 \\
        PCA & contemp., & ranks & 2 & unio & yes & 5 & 0.025 & 0.0164 & 2.3868 & 1.3622 \\
        & $\text{salesGR}_{1}$, &  &  &  &  &  &  &  \\
        & $\text{opmar}\Delta_{1}$ &  &  &  &  &  &  &  \\
        RD & $\beta$, P/E & -- & -- & union & no & 10 & 0.05 & 0.0213 & 3.9850 & 4.9765 \\
        PCA & sales, at, seq, & $x^{1/5}$ & 90\% & union & no & 10 & 0.025 & 0.0217 & 2.6242 & 2.4904 \\
        & P/E, P/B &  &  &  &  &  &  &  \\
        PCA & sales, opmar, & trim & 2/ & union & yes & 5 & 0.025 & 0.0233 & 5.6027 & 7.1741 \\
        &  at, seq &  & 75\% &  &  &  &  &  \\
        single & opmar & -- & -- & -- & -- & 10 & 0.05 & 0.0284 & 4.1500 & 6.1454 \\
        single & opmar & -- & -- & -- & -- & 5 & 0.05 & 0.0309 & 4.4533 & 4.8490 \\
        market & -- & -- & -- & -- & -- & 5 & -- & 0.0454 & 6.0073 & 11.1804 \\
        MC & sales & -- & -- & -- & -- & 5 & -- & 0.0516 & 6.3825 & 12.7518 \\
        group & major group & -- & -- & -- & -- & 5 & -- & 0.0653 & 8.6576 & 22.5482 \\
    \end{tabular}
\end{table}

Across all forecast horizons the algorithms using several reference variables improve distributional forecast performance reducing $\Delta_{\text{q}}$ by between 38\% and 71\%.
Generally, the reduction is greater with shorter forecast horizons and overall the results improve with a shorter forecast horizon.
For all forecast horizons the best results are delivered by PCARDs with a fixed number of PCs based on contemporaneous reference variables combined with different degrees of lagged \textit{past operating margin deltas} and \textit{past sales growth rates}.
However, past reference variables do not exceed a lag of five years.
All reference variables should be either 2.5\% trimmed on both tails before PCA or one should use ranks for PCA.
Overall a window length of 30 years is best, besides a 20-year window for three-year horizon, while combining reference variables by union.
The choices of reference class size and correction vary among the best results. 
Under some constraints, LARD and intersection are among the best combination versions and the number of PCs gets chosen by a data driven criterion although the overall best results still use a fixed number of PCs and union for combination.

All contemporaneous variables are used for the best combination and under certain constraints the market parameters $\beta$, P/E and P/B are important.
An exception is the 10-year horizon where only balance sheet variables are selected with a possible interpretation that market parameters better reflect short term expectations.
The concentrated information in \textit{past operating margin deltas} reflects a company's market position and, thus, it is reasonable to assume a positive influence of growing operating margin on future sales growth.
For a discussion on reasons thereof, compare \cite{theisingetal2023} and the literature mentioned there.
The uncompetitive performance of \textit{sales} may be due to Gibrat's law that states that firm growth is independent of firm size \citep{gibrat1931} but the excellent performance of \textit{past sales growth rates} contradicts the part of Gibrat's law that claims growth rates are uncorrelated in time.
\cite{stanleyetal96} also show that sales growth depends on past growth rates and that sales growth distributions are similar across diverse firms which corresponds to the poor results of the group approach here.

Using multiple reference variables with RD improves the results by between 58\% and 12\% compared to the single variable use.
The \textit{past operating margin deltas} and \textit{past sales growth rates} dominate the forward selection with lags mainly between three and eight years, partially up to 10 years, a window length of 30 years and the number of used reference variables in the best combination varies from two to five (c.f. Tables \ref{tab:resh1forward}, \ref{tab:resh3forward}, \ref{tab:resh5forward} and \ref{tab:resh10forward} in Appendix \ref{app:refvartables}).
Taking more reference variables into consideration does not yield better results in general which seems to be a feature of the specific rank based algorithms used here, e.g., combining contemporaneous variables with past sales growth rates and past operating margin up to different lags performs substantially worse compared to single reference variables with $\Delta_{\text{q}}$ between $1.7$ and $3.7$ times higher depending on forecast horizon (see Table \ref{tab:resrdcont} in Appendix \ref{app:refvartables}).
However, this may partially explain the outstanding performance of PCARD using a small fixed number of PCs.
As for PCARD, the best RD options vary across size and correction but all combine the reference variables by union.
To put this into perspective, union means that the reference class members are similar to the initial firm in at least one reference variable in contrast to similarity across variables when using LARD or intersection.
This superiority reflects the fact that the outside view, often seen as relying on \textit{only superficially similar instances}, produces more accurate predictions than a narrow-minded focus on the uniqueness and complexity of a forecast challenge \citep[c.f.][]{loka2003}.
The benchmarks of market climate and group approach as well as of \cite{maubcal15} are clearly outperformed across all forecast horizons.
There are some different rankings of algorithms across accuracy measures which is natural as there is no universal ranking of forecasts regardless of the accuracy measure \citep[see][]{dieboldetal1998} and given the focus of KS and CvM on the complete distribution.

Considering the best results for forecasting 1-year sales growth from Table \ref{tab:resh1rd} we put the accuracy measure $\Delta_{\text{q}}$ into perspective.
Here, we have the sum of nine absolute quantile deviations $\Delta_{\text{q}} = 0.0045$ which means that according to the backtest we miss the quantile levels of the distribution of one-year sales growth on average by 0.05 percentage points when predicting on historical data.
If a practioner, e.g., derives 90\% confidence intervals based on a reference class selection with the best algorithm options this average error should be negligible.

\begin{table}
    \center
    \caption{Comparison of reference variables (ref. var.) and algorithms for forecasting three-years ahead sales growth. Alg. is algorithm, transf. is pre PCA transformation, comb. is combination of reference variables, cor. is correction and inters. is intersection.}
    \label{tab:resh3rd}
    \footnotesize
    \begin{tabular}{llllllll|rrr}
        Alg. & Ref. Var. & Transf. & \#{}PC & Comb. & Cor. & $w$ & Size & $\Delta_{\text{q}}$ & KS & CvM \\
        \midrule
        PCA & contemp., & trim & 3 & union & yes & 20 & 0.01 & 0.0146 & 2.2379 & 1.3913 \\
        & $\text{salesGR}_{1:3}$,  &  &  &  &  &  &  &  \\
        & $\text{opmar}\Delta_{1:3}$ &  &  &  &  &  &  &  \\
        PCA & contemp., & trim & 3 & union & yes & 20 & 0.01 & 0.0163 & 1.9873 & 1.1432 \\
        & $\text{salesGR}_{1:5}$,  &  &  &  &  &  &  &  \\
        & $\text{opmar}\Delta_{1:5}$ &  &  &  &  &  &  &  \\
        PCA & contemp., & trim & 3 & union & no & 20 & 0.05 & 0.0165 & 2.2956 & 1.4306 \\
        & $\text{salesGR}_{1:5}$,  &  &  &  &  &  &  &  \\
        & $\text{opmar}\Delta_{1:5}$ &  &  &  &  &  &  &  \\
        RD & $\text{salesGR}_{10}$, & -- & -- & union & no & 30 & 0.01 & 0.0223 & 1.7135 & 0.7237 \\
        & $\text{opmar}\Delta_{8,9}$ &  &  &  &  &  &  &  \\
        RD & $\text{salesGR}_{10}$, & -- & -- & union & yes & 30 & 0.01 & 0.0233 & 1.7000 & 0.6886 \\
        & $\text{opmar}\Delta_{8,10}$ &  &  &  &  &  &  &  \\
        PCA & sales, opmar, & trim & $\mathrm{Var}_{\mu}$ & union & yes & 30 & 0.01 & 0.0238 & 3.1099 & 2.3311 \\
        &  seq, $\beta$, &  &  &  &  &  &  &  \\  
        &  P/E, P/B &  &  &  &  &  &  &  \\
        RD & $\text{salesGR}_{10}$, & -- & -- & union & yes & 30 & 0.01 & 0.0239 & 1.4898 & 0.5776 \\
        & $\text{opmar}\Delta_{8}$ &  &  &  &  &  &  &  \\
        PCA & $\text{opmar}\Delta_{6:8,10}$ & -- & 75\% & inters. & no & 10 & 0.01 & 0.0282 & 1.3639 & 0.3202 \\
        single & $\text{opmar}\Delta_{7}$ & -- & -- & -- & -- & 30 & 0.025 & 0.0290 & 3.2390 & 2.8895 \\
        RD & $\beta$, P/E, P/B & -- & -- & LARD & -- & 10 & 0.05 & 0.0319 & 4.7188 & 6.6191 \\
        RD & $\beta$, P/E, P/B & -- & -- & LARD & -- & 5 & 0.025 & 0.0334 & 4.0785 & 3.7155 \\
        PCA & $\text{opmar}\Delta_{6:10}$ & -- & 75\% & inters. & no & 5 & 0.025 & 0.0373 & 3.0515 & 2.6985 \\
        PCA & sales, opmar, & $x^{1/5}$ & 2 & union & yes & 5 & 0.05 & 0.0460 & 6.6272 & 12.6316 \\
        & $\beta$, P/E &  &  &  &  &  &  &  \\
        PCA & sales, opmar, & $x^{1/5}$ & 2 & union & yes & 10 & 0.05 & 0.0486 & 6.5503 & 12.6209 \\
        & seq, $\beta$, &  &  &  &  &  &  &  \\
        & P/E, P/B &  &  &  &  &  &  &  \\
        single & opmar & -- & -- & -- & -- & 30 & 0.01 & 0.0603 & 6.9167 & 16.7652 \\
        single & opmar & -- & -- & -- & -- & 5 & 0.05 & 0.0707 & 10.4687 & 33.6970 \\
        single & opmar & -- & -- & -- & -- & 10 & 0.05 & 0.0883 & 11.8219 & 55.6199 \\
        market & -- & -- & -- & -- & -- & 5 & -- & 0.0924 & 11.3359 & 45.3895 \\
        MC & sales & -- & -- & -- & -- & 5 & -- & 0.1028 & 13.4856 & 61.3178 \\
        group & major group & -- & -- & -- & -- & 5 & -- & 0.1423 & 17.9423 & 106.9768 \\
    \end{tabular}
\end{table}

With practical application in mind the amount of necessary data is important and consists of two components, the years used to select a window of candidates and the lags of past sales growth and operating margin delta.
Collecting a smaller data basis is easier to achieve in practice and further takes into account that practitioners might want to assume a stable data generating mechanism of sales growth for only a few years.
However, smaller windows generate smaller candidate sets and ultimately smaller reference classes in general.
In particular, past operating margin deltas and past sales growth rates turn out to be best in the forward selection but depend on a rather large amount of data.
The best performances by PCARDs need between 23 and 35 years of data while the best RD and the best single variable combinations need between 36 and 40 years of data despite performing worse.
Remarkably, the use of dimension reduction via PCA enables us to incorporate more reference variables while simultaneously needing less data and improving the results substantially.

\begin{table}
    \center
    \caption{Comparison of reference variables (ref. var.) and algorithms for forecasting five-years ahead sales growth. Alg. is algorithm, transf. is pre PCA transformation, comb. is combination of reference variables, cor. is correction and inters. is intersection.}
    \label{tab:resh5rd}
    \footnotesize
    \begin{tabular}{llllllll|rrr}
        Alg. & Ref. Var. & Transf. & \#{}PC & Comb. & Cor. & $w$ & Size & $\Delta_{\text{q}}$ & KS & CvM \\
        \midrule
        PCA & contemp., & trim & 2 & union & yes & 30 & 0.01 & 0.0179 & 1.0948 & 0.2297 \\
        & $\text{salesGR}_{1:5}$, &  &  &  &  &  &  &  \\
        & $\text{opmar}\Delta_{1:5}$ &  &  &  &  &  &  &  \\
        PCA & contemp., & trim & 2 & LARD & -- & 30 & 0.01 & 0.0186 & 1.5236 & 0.8604 \\
        & $\text{salesGR}_{1:5}$, &  &  &  &  &  &  &  \\
        & $\text{opmar}\Delta_{1:5}$ &  &  &  &  &  &  &  \\
        PCA & contemp., & trim & 2 & union & yes & 30 & 0.025 & 0.0207 & 1.1641 & 0.3481 \\
        & $\text{salesGR}_{1:5}$, &  &  &  &  &  &  &  \\
        & $\text{opmar}\Delta_{1:5}$ &  &  &  &  &  &  &  \\
        RD & $\text{salesGR}_{10}$, & -- & -- & union & yes & 30 & 0.01 & 0.0230 & 2.2279 & 1.0487 \\
        & $\text{opmar}\Delta_{6}$ &  &  &  &  &  &  &  \\
        RD & $\text{salesGR}_{10}$, & -- & -- & union & no & 30 & 0.01 & 0.0261 & 2.1110 & 1.0450 \\
        & $\text{opmar}\Delta_{6}$ &  &  &  &  &  &  &  \\
        PCA & sales, opmar, & trim & 90\% & union & no & 30 & 0.05 & 0.0261 & 3.0871 & 1.7590 \\
         & seq, $\beta$, &  &  &  &  &  &  &  \\
         & P/E, P/B &  &  &  &  &  &  &  \\
        RD & $\text{salesGR}_{6, 10}$, & -- & -- & union & yes & 30 & 0.01 & 0.0264 & 1.5280 & 0.6581 \\
        & $\text{opmar}\Delta_{6}$ &  &  &  &  &  &  &  \\
        single & $\text{opmar}\Delta_{10}$ & -- & -- & -- & -- & 30 & 0.01 & 0.0320 & 2.2045 & 1.3087 \\
        PCA & sales, opmar, & trim & $\mathrm{Var}_{\mu}$ & inters. & no & 5 & 0.025 & 0.0394 & 1.3993 & 0.1071 \\
        & at, seq, &  &  &  &  &  &  &  \\
        & $\text{salesGR}_{1:5}$ &  &  &  &  &  &  &  \\
        & $\text{opmar}\Delta_{1:5}$ &  &  &  &  &  &  &  \\
        PCA & $\text{opmar}\Delta_{4:7,9,10}$ & -- & $\mathrm{Var}_{\mu}$ & inters. & no & 10 & 0.025 & 0.0399 & 4.2461 & 5.1859 \\
        PCA & sales, opmar, & trim & 75\% & inters. & yes & 5 & 0.025 & 0.0484 & 1.2470 & 0.1948 \\
         & $\beta$, P/E, P/B &  &  &  &  &  &  &  \\
        RD & $\beta$, P/E & -- & -- & LARD & -- & 30 & 0.05 & 0.0492 & 5.1695 & 6.0082 \\
        PCA & sales, opmar, & trim & 2 & inters. & no & 10 & 0.01 & 0.0559 & 1.7647 & 0.3353 \\
         & $\beta$, P/E, P/B &  &  &  &  &  &  &  \\
        RD & opmar,  & -- & -- & inters. & yes & 5 & 0.01 & 0.0634 & 2.1629 & 0.4439 \\
        & $\text{opmar}\Delta_{6}$ &  &  &  &  &  &  &  \\
        RD & opmar, & -- & -- & inters. & yes & 10 & 0.01 & 0.0794 & 2.4536 & 0.3912 \\
        & $\text{opmar}\Delta_{10}$ &  &  &  &  &  &  &  \\
        RD & $\beta$, P/E, P/B & -- & -- & LARD & -- & 5 & 0.05 & 0.0716 & 6.7605 & 12.5422 \\
        single & opmar & -- & -- & -- & -- & 30 & 0.01 & 0.0856 & 9.4768 & 32.0558 \\
        RD & opmar, P/B & -- & -- & inters. & no & 10 & 0.05 & 0.0863 & 1.3772 & 0.3369 \\
        single & P/E & -- & -- & -- & -- & 5 & 0.05 & 0.1113 & 9.1733 & 41.2639 \\
        market & -- & -- & -- & -- & -- & 5 & -- & 0.1428 & 15.2365 & 96.8583 \\
        single & P/E & -- & -- & -- & -- & 10 & 0.05 & 0.1497 & 13.1570 & 84.6791 \\
        MC & sales & -- & -- & -- & -- & 5 & -- & 0.1600 & 19.0380 & 137.3940 \\
        group & major group & -- & -- & -- & -- & 30 & -- & 0.2136 & 16.7058 & 106.9918 \\
    \end{tabular}
\end{table}

\begin{table}
    \center
    \caption{Comparison of reference variables (ref. var.) and algorithms for forecasting ten-years ahead sales growth. Alg. is algorithm, transf. is pre PCA transformation, comb. is combination of reference variables, cor. is correction and inters. is intersection.}
    \label{tab:resh10rd}
    \footnotesize
    \begin{tabular}{llllllll|rrr}
         Alg. & Ref. Var. & Transf. & \#{}PC & Comb. & Cor. & $w$ & Size & $\Delta_{\text{q}}$ & KS & CvM \\
        \midrule
        PCA & sales, opmar, & trim & 2 & union & yes & 30 & 0.01 & 0.0275 & 1.5735 & 0.6993 \\
        & at, seq, &  &  &  &  &  &  &  \\
        & $\text{salesGR}_{1:5}$ &  &  &  &  &  &  &  \\
        & $\text{opmar}\Delta_{1:5}$ &  &  &  &  &  &  &  \\
        PCA & sales, opmar, & trim & 2 & LARD & -- & 30 & 0.01 & 0.0284 & 1.7164 & 0.8456 \\
        & at, seq, &  &  &  &  &  &  &  \\
        & $\text{salesGR}_{1:5}$ &  &  &  &  &  &  &  \\
        & $\text{opmar}\Delta_{1:5}$ &  &  &  &  &  &  &  \\
        PCA & sales, opmar, & trim & 2 & union & yes & 30 & 0.025 & 0.0296 & 1.9586 & 0.7830 \\
        & at, seq, &  &  &  &  &  &  &  \\
        & $\text{salesGR}_{1:5}$ &  &  &  &  &  &  &  \\
        & $\text{opmar}\Delta_{1:5}$ &  &  &  &  &  &  &  \\
        PCA & sales, opmar, & trim & $\mathrm{Var}_{\mu}$ & inters. & no & 20 & 0.01 & 0.0323 & 1.5376 & 0.4871 \\
        & at, seq, $\beta$ &  &  &  &  &  &  &  \\
        RD & $\text{salesGR}_{6, 7}$, & -- & -- & union & yes & 30 & 0.01 & 0.0388 & 3.0820 & 3.2019 \\
        & $\text{opmar}\Delta_{5, 7}$ &  &  &  &  &  &  &  \\
        RD & $\text{salesGR}_{6}$, & -- & -- & union & yes & 30 & 0.01 & 0.0401 & 3.1193 & 3.2598 \\
        & $\text{opmar}\Delta_{7}$ &  &  &  &  &  &  &  \\
        RD & $\text{salesGR}_{6:8}$, & -- & -- & union & yes & 30 & 0.01 & 0.0403 & 2.9479 & 2.9308 \\
        & $\text{opmar}\Delta_{5, 7, 8}$ &  &  &  &  &  &  &  \\
        single & $\text{opmar}\Delta_{6}$ & -- & -- & -- & -- & 30 & 0.025 & 0.0441 & 3.7773 & 4.1454 \\
        PCA & opmar, $\beta$, & trim & 90\% & inters. & no & 5 & 0.05 & 0.0584 & 1.1665 & 0.2001 \\
        & P/E, P/B  &  &  &  &  &  &  &  \\
        RD & $\beta$, P/E & -- & -- & LARD & -- & 30 & 0.025 & 0.0679 & 4.1868 & 4.2020 \\
        PCA & opmar, $\beta$, & trim & 3 & inters. & yes & 10 & 0.01 & 0.0705 & 1.2347 & 0.4220 \\
        & P/E, P/B  &  &  &  &  &  &  &  \\
        RD & $\beta$, $\text{salesGR}_{5}$, & -- & -- & LARD & -- & 10 & 0.05 & 0.0785 & 6.0266 & 10.9679 \\
        & $\text{opmar}\Delta_{7}$  &  &  &  &  &  &  &  \\
        RD & $\text{salesGR}_{7}$, & -- & -- & inters. & no & 5 & 0.025 & 0.0998 & 1.0023 & 0.1382 \\
        & $\text{opmar}\Delta_{3, 5, 6}$ &  &  &  &  &  &  &  \\
        single & opmar & -- & -- & -- & -- & 30 & 0.01 & 0.1126 & 7.4249 & 20.6290 \\
        RD & opmar, $\beta$, P/E & -- & -- & LARD & -- & 10 & 0.05 & 0.1147 & 8.6451 & 17.1154 \\
        RD & opmar, $\beta$, P/E & -- & -- & LARD & -- & 5 & 0.05 & 0.1154 & 9.5327 & 24.3167 \\
        single & $\text{opmar}\Delta_{10}$ & -- & -- & -- & -- & 10 & 0.025 & 0.2053 & 8.5536 & 31.8502 \\
        MC(15) & sales & -- & -- & -- & -- & 30 & -- & 0.2270 & 11.2416 & 50.6546 \\
        group & major group & -- & -- & -- & -- & 30 & -- & 0.2561 & 12.0198 & 61.4773 \\
        market & -- & -- & -- & -- & -- & 30 & -- & 0.2845 & 14.4029 & 74.5287 \\
        single & $\text{opmar}\Delta_{10}$ & -- & -- & -- & -- & 5 & 0.025 & 0.2926 & 14.5939 & 94.3775 \\
    \end{tabular}
\end{table}

Further, the following results from Tables \ref{tab:resh1rd} - \ref{tab:resh10rd} stand out, where changes are reported with respect to the best single reference variable for the respective forecast horizon.
For one-year forecast horizon (see Table \ref{tab:resh1rd}) PCARD improves $\Delta_{\text{q}}$ by 71\% and RD improves $\Delta_{\text{q}}$ by 58\% while using five years less and one year more of data, respectively.
There is even a 39\% improvement for PCARD with 25 years less of necessary data.
Notably, RD using $\beta$ and P/E performs roughly the same as the best single reference variable but needs only five instead of 36 years of data.
Interestingly, for the three best PCARD combinations the results for three PCs and all PCs that explain at least the mean variance are identical.
When forecasting three years ahead, Table \ref{tab:resh3rd} shows a reduction of $\Delta_{\text{q}}$ by 50\% and by 23\% for PCARD and RD with 14 years less and three years more of data, respectively.
Even PCARD with using 17 years less of information is slightly better and additionally, PCARD with only contemporaneous variables and a 30-year window is better than the single variable method as well.
Moreover, findings on five-year ahead forecasting in Table \ref{tab:resh5rd} show an improvement of $\Delta_{\text{q}}$ by 44\% and by 28\% for PCARD and RD while needing five years less and the same amount of past data.
In contrast, a PCARD combination with contemporaneous variables and a 30-year window improves results by 18\% while needing 10 years less of data.
Finally, in Table \ref{tab:resh10rd} on 10-year forecast horizon it stands out that contemporaneous balance sheet variables and past variables are the best reference variables for PCARD and reduce $\Delta_{\text{q}}$ by 38\% with similar necessary data.
While the best RD combinations have a comparable data demand as the single variable approach and improve $\Delta_{\text{q}}$ by 12\%, notably, a PCA combination using intersection on only contemporaneous variables from a 20-year window improves results by 27\% despite using 16 years less of information.

\section{Practical Application}

In an actual forecast challenge in any field of application the resulting information in terms of a distributional forecast can be used in several ways, from assessing an existing forecast over providing confidence intervals and point estimates to calculating moments and probabilities for intervals of possible outcomes. 
By means of the resulting information we can assess predictions (e.g. model based or by experts, analysts) by evaluating the ECDF of the reference class at the prediction, i.e. we calculate the PIT value $\mathbbm{P} (Y_{i,t+h} \leq y_{i,t+h}) \approx n^{-1} \sum_{(j,s)\in R} \mathbbm{1}\{ Y_{j,s+h} \leq y_{i,t+h}\}$ for $n=\vert R\vert$. 
PIT values very close to either 0 or 1 can serve as a warning signal to revisit a prediction and check for arguments that may justify the prediction relative to the reference class or possibly correct the prediction.
Confidence intervals and point estimates can be directly calculated from the ECDF as well as probabilities for intervals of possible outcomes.
Here, we provide an application of the reference class approach on forecasting sales growth for firms over multiple years and additionally assess expert forecasts.

\begin{figure}
	\center
	\includegraphics[height=0.33\textheight]{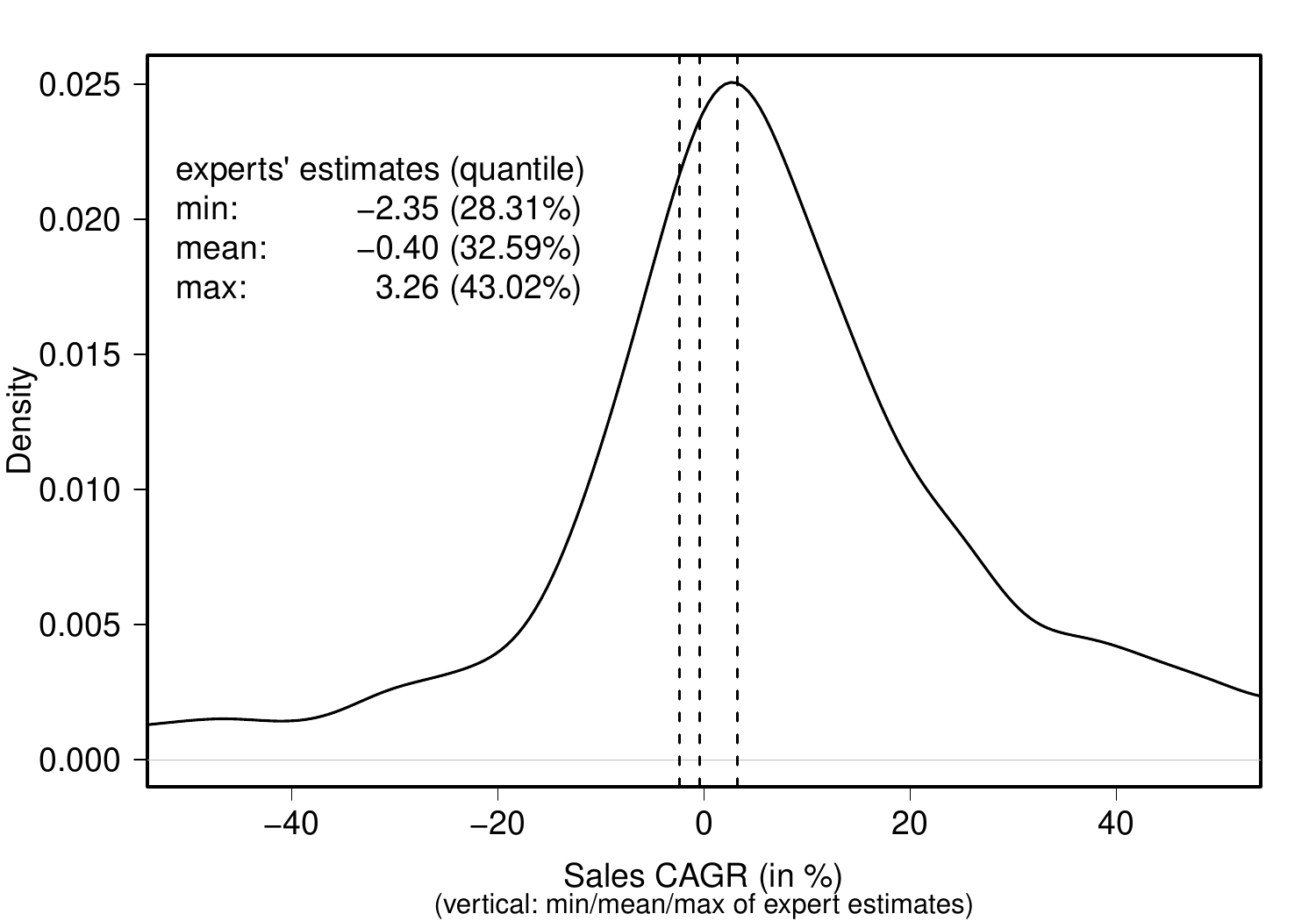}
	\caption{Forecasted density of one-year sales growth for 3M based on the best algorithm options from Table \ref{tab:resh1rd} compared to experts' estimates. Density estimation on support $[-100,\infty )$ is based on the Gaussian kernel and Silverman's rule of thumb provdies the bandwidth.}
	\label{fig:example-dreim-1}
	\includegraphics[height=0.33\textheight]{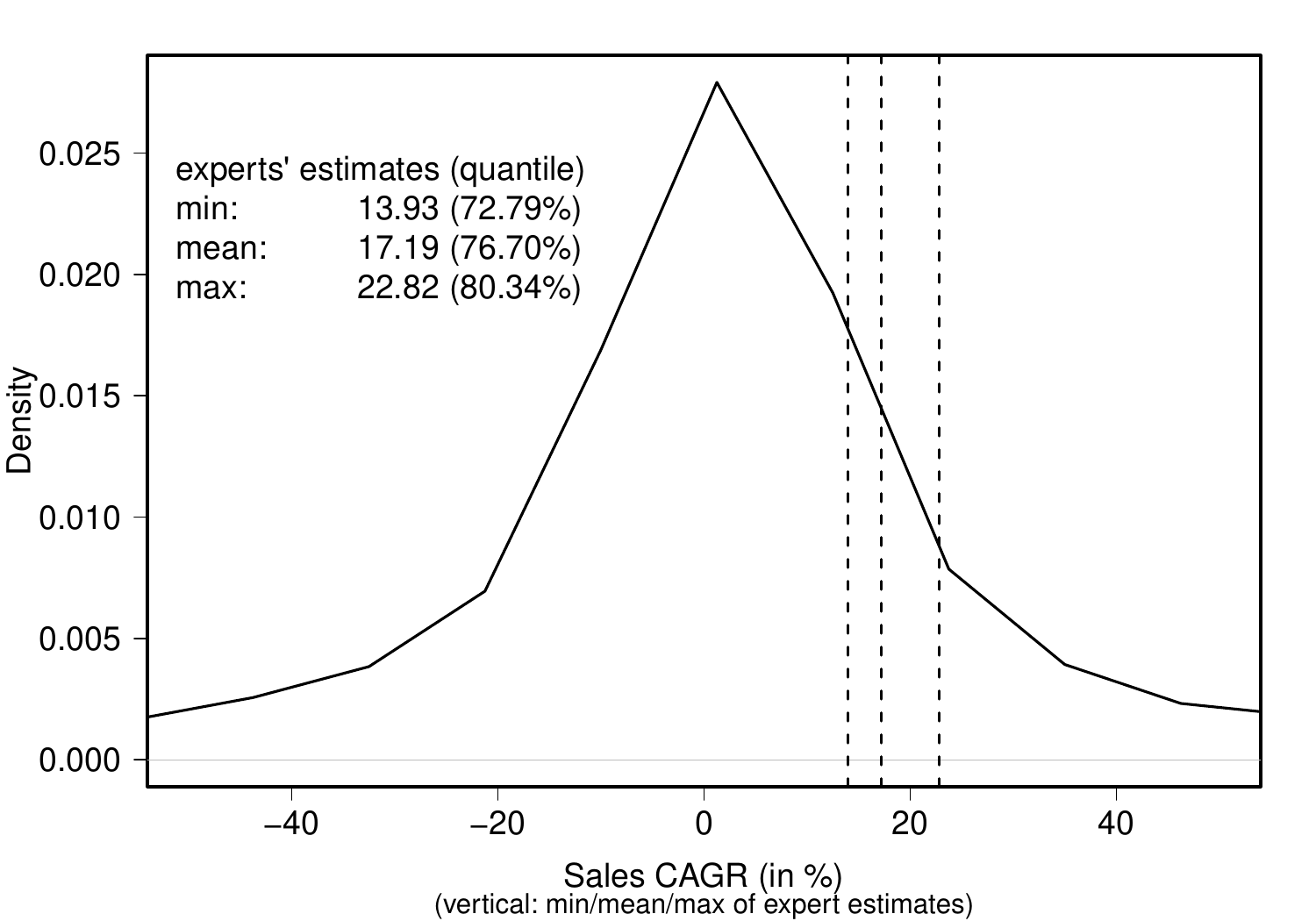}
	\caption{Forecasted density of one-year sales growth for Amazon based the best algorithm options from Table \ref{tab:resh1rd} compared to experts' estimates. Density estimation on support $[-100,\infty )$ is based on the Gaussian kernel and Silverman's rule of thumb provdies the bandwidth.}
	\label{fig:example-amazon-1}
\end{figure}

Comparing expert forecasts of corporate sales growth to base rates displays the practical utility of reference classes.
We compare the distributional forecasts based on the reference classes to expert forecasts which infamously tend to neglect available data of similar cases.
As in \cite{theisingetal2023}, we discuss two firms as an example of reference class selection -- 3M and Amazon.
For both companies, we forecast the distribution of one-year sales growth rate based on a reference class.
The distributional forecasts are then compared to analysts' forecast from the \cite{factset} estimates database.
For both forecasts $2018$ is the base year and Figures \ref{fig:example-dreim-1} and \ref{fig:example-amazon-1} display the results.
The reference classes for one-year sales growth forecasts in this section are selected according to the best result from our backtest in Table \ref{tab:resh1rd} using all contemporaneous variables, one-year past sales growth and one-year operating margin delta as reference variables.
The Selection is based on a PCA rotation of the reference variables' ranks with 3 PCs where candidates are chosen from a 30 year window period and the reference classes for the individual PCs have the size $0.33\%$ of the candidate set due to correction and the three reference classes are then unified.

Sales growth of 3M was predicted 15 times and Figure \ref{fig:example-dreim-1} shows that the expert forecasts lie between -2.35\% and 3.26\% sales growth and range from the 28.31\% to the 43.02\% quantile within the reference class.
This indicates no sign of overoptimism as the expert forecasts are close to the center of the distributional forecast.
Thus, inside and outside view roughly agree and classify 3M as an average company in terms of sales growth.
However, a low range of expert forecasts in comparision to the reference class forecast is emphasized by a predicted coverage rate of only 14.71\% that could lead to overconfidence in the interval of expert forecasts.
To put the size of the reference class into context, there are 271,548 firm observations before 2018 available in our data set and restricting them to all firms providing the necessary reference variables in a 30 year window shrinks these to a set of 109,792 candidate firms.
Out of these candidates, the algorithm chooses 1,074 observations or 0.98\% of the candidates.
To put this into perspective, the best combination with a window 10-year in Table \ref{tab:resh1rd}, a set of options with 20 years less data consumption, operates on a set of 27,346 candidates and selects only 541 of them, that is 1.98\%, as candidates.
Note that the former algorithm option used a size correction and the latter not, resulting in a reference class twice as big as the size parameter suggested as there are few observations that are chosen according to both PCs.

For Amazon, there are 43 expert forecasts and Figure \ref{fig:example-amazon-1} compares them to the distribution within Amazon's reference class.
The forecasts vary more than for 3M, namely from 13.93\% to 22.82\%, and their relation to the reference class is different.
Here, forecasts correspond to quantiles between 72.79\% and 80.34\% and are more optimistic since only one out of five firms in the reference class achieved the maximum predicted growth of Amazon.
The difference between inside and outside should prompt forecasters to reconsider and question their predictions.
Amazon is known for its capability of high growth, however, there should be valid reasons for an optimistic forecast with respect to the base rates.
The outside view may at least protect against extreme and unrealistic predictions.
Again, we discuss the actual size of the reference class, where the set of candidate firms is the same as for 3M because forecast horizon, base year and algorithm options are identical.
The reference class then consists of 1,073 firms, i.e. 0.98\% of the candidates, as opposed to selecting 542 firms or 1.98\% of the candidates for the less data consuming algorithm as for 3M using a 10-year window.

\begin{table}
    \center
    \caption{Comparision of reference classes for forecasting compound annual sales growth rates of 3M and Amazon with base year 2018. The choice of algorithm for each forecast horizon is based on the results from the backtests in Tables \ref{tab:resh1rd} -- \ref{tab:resh10rd}. Mean and standard deviation are 2.5\% trimmed on both tails.}
    \label{tab:refclassdreimamazon}
    \begin{tabular}{l|rr|rr|rr|rr}
        & \multicolumn{8}{c}{Base Rates}\\
        \midrule
        & \multicolumn{2}{c}{1-Yr} & \multicolumn{2}{c}{3-Yr} & \multicolumn{2}{c}{5-Yr} & \multicolumn{2}{c}{10-Yr} \\
        CAGR (\%) & 3M & Amazon & 3M & Amazon & 3M & Amazon & 3M & Amazon \\
        \midrule
        $\leq -25$ & 7.54 & 10.81 & 2.70 & 5.45 & 1.51 & 2.57 & 0.60 & 1.40 \\
        $]-25,-20]$ & 2.05 & 2.52 & 1.98 & 1.64 & 1.06 & 0.76 & 0.20 & 0.40 \\
        $]-20,-15]$ & 1.58 & 3.45 & 1.98 & 2.73 & 2.57 & 1.81 & 1.20 & 1.00 \\
        $]-15,-10]$ & 4.00 & 3.45 & 5.41 & 2.73 & 3.78 & 4.23 & 2.80 & 3.00 \\
        $]-10,-5]$ & 7.45 & 7.83 & 9.19 & 7.27 & 9.23 & 5.74 & 9.00 & 7.00 \\
        $]-5,0]$ & 11.17 & 13.05 & 17.84 & 11.45 & 19.97 & 14.50 & 23.80 & 18.80 \\
        $]0,5]$ & 14.25 & 16.50 & 20.36 & 18.55 & 28.74 & 21.75 & 38.20 & 27.00 \\
        $]5,10]$ & 10.89 & 9.23 & 13.15 & 13.09 & 16.34 & 15.26 & 14.00 & 14.80 \\
        $]10,15]$ & 8.94 & 7.08 & 9.73 & 10.18 & 8.17 & 12.69 & 6.00 & 11.80 \\
        $]15,20]$ & 6.15 & 4.75 & 6.67 & 6.91 & 4.39 & 6.50 & 2.40 & 5.40 \\
        $]20,25]$ & 4.56 & 2.61 & 3.24 & 4.18 & 2.12 & 4.68 & 0.60 & 5.20 \\
        $]25,30]$ & 3.91 & 2.52 & 0.90 & 3.27 & 1.06 & 3.47 & 0.60 & 1.80 \\
        $]30,35]$ & 1.86 & 1.49 & 1.44 & 3.27 & 0.45 & 2.72 & 0.40 & 1.00 \\
        $]35,40]$ & 2.33 & 1.21 & 1.44 & 2.00 & 0.30 & 0.60 & 0.20 & 0.40 \\
        $]40,45]$ & 1.86 & 1.03 & 1.26 & 2.00 & 0.15 & 0.45 & 0.00 & 0.40 \\
        $>45$ & 11.45 & 12.49 & 2.70 & 5.27 & 0.15 & 2.27 & 0.00 & 0.60 \\
        \midrule
        mean & 10.63 & 11.98 & 4.13 & 7.16 & 2.23 & 6.16 & 1.63 & 4.57 \\
        \midrule
        median & 5.99 & 2.73 & 2.81 & 5.08 & 2.36 & 4.65 & 1.38 & 3.21 \\
        \midrule
        std & 26.40 & 47.85 & 12.57 & 16.95 & 8.09 & 11.55 & 5.85 & 8.75 \\
        \midrule
    \end{tabular}
\end{table}

Table \ref{tab:refclassdreimamazon} shows base rates for one, three, five and 10 year forecasting horizons for 3M and Amazon and underlines the higher growth chances of Amazon.
We demonstrate the usefulness of these base rates by considering, e.g., an entity that wants to invest their money in rising buisnesses that have the highest probability of a long term sales growth above 5\% per year.
Assuming 10 years to be long term, we can directly use the base rates in Table \ref{tab:refclassdreimamazon} to predict such a probability for both companies by adding up the relevant cells in the most right columns.
That results in predicted probabilities of 24.2\% and 41.4\% for more than 5\% compound annual sales growth for 3M and Amazon, respectively.
Especially the base rates for larger sales growth  are higher for Amazon than for 3M, for all forecast horizons and, in general, the predicted distribution has a higher variability for the former.
This may be interpreted as the higher risk that contemplates the higher potential reward.
Overall, the standard deviation declines with the forecast horizon as we consider compound annual growth rates.
Interestingly, for one-year forecast horizon the base rates for 3M show less probability for sales decline than for Amazon and, except for growth above 45\%, 3M has higher base rates for sales increase.
This contemplates Figure \ref{fig:example-dreim-1} where the expert forecasts are roughly centered at unchanged sales but the distributional reference class forecast shows a tendency to sales increase.
However, the base rates for one-year sales growth exhibit by far the most uncertainty, as can be seen by the trimmed standard deviation, and also the highest predicted probabilites for sales growth exceeding 45\% and more than 25\% probability of sales decline.

\begin{figure}
	\center
	\includegraphics[height=0.33\textheight]{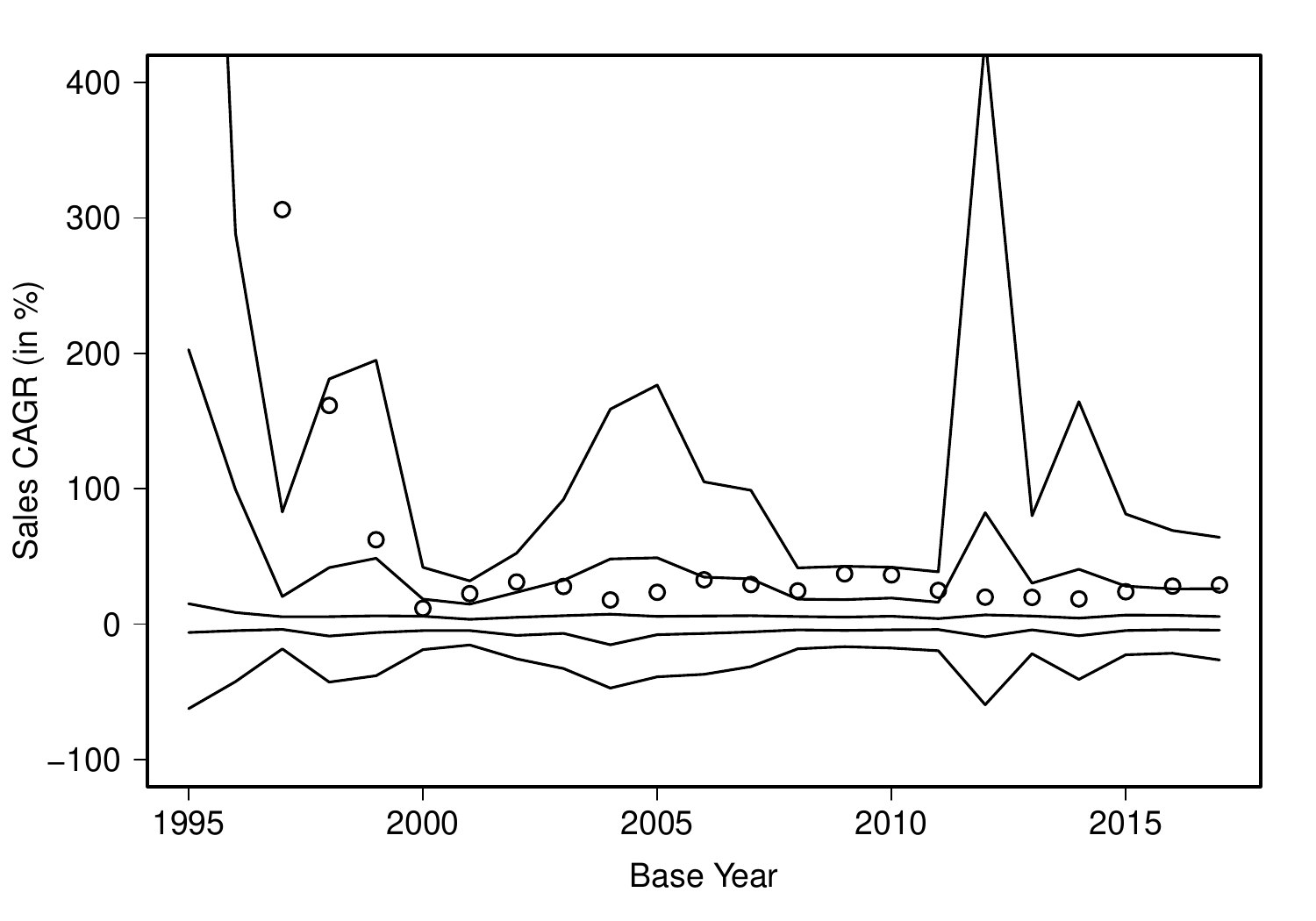}
	\caption{One year sales growth for Amazon from 1995 to 2017 compared to quantiles of the reference class outcomes selected using the best algorithm options from Table \ref{tab:resh1rd}. The bold lines from bottom to top represent the 10\%, 25\%, 50\%, 75\% and 90\% quantiles of one year sales growth within the reference class. The circles represent sales growth of Amazon which are 2,891\% and 823\% for base years 1995 and 1996, respectively, and therefore not displayed in this graph.}
	\label{fig:example-amazon-1-historic}
\end{figure}

Moreover, we stick to the example of Amazon and present another useful application of reference classes.
Figure \ref{fig:example-amazon-1-historic} follows Amazon's one year ahead distibutional sales growth forecasts through the base years 1995 -- 2017 and compares these forecasts to the realized sales growth.
Here, it shows that Amazon is outperforming its reference class massively in the first three years.
After that, the realized sales growth rate is close to the 75\% quantile for most of the years, with some fluctuations up and down between the 50\% and 90\% quantile of its reference classes.
With respect to the situation in Figure \ref{fig:example-amazon-1} this may serve as a validation of expert forecasts.
More general, the dependence of reference class selection on historic data gets evident as the distributional forecasts' uncertainty increases in the aftermath of well-known times of financial distress, here, the dotcom crisis in 2000, the subprime bubble in 2007 and 2008, and the European debt crisis in 2009 and 2010.
Overall, a practioner might conclude Amazon performs well compared to peers and is in a good overall market position.
On the one hand, this seems like old news, but on the other hand it serves as an affirmation and a proof of concept:
Reference class forecasting is well behaved and meets practical expectations in this case.

\section{Concluding Remarks}

In this paper, we extend the analysis of distributional reference class forecasting of corporate sales growth with a focus on reference class selection.
We provide a practical solution to the well-known reference class problem \citep{venn1888} that arises in any application of reference class forecasting as desribed in \cite{kahnemantversky79}.
The novel rank based methods allow several reference variables and include the option of a dimension reduction based on principal components.
In an extensive backtest on corporate data from the USA covering several decades we conclude that especially principal component analysis reduces the amount of necessary past data while simultaneously improving previous approaches \citep{theisingetal2023} substantially by between 38\% and 71\% depending on forecast horizon.
Further, we illustrate the practical usefulness of the new methods on two example firms, 3M and Amazon.
The novel approaches need less historic observations compared to existing methods, are easy to interpret and deliver reasonable results making them useful for practical applications.
However, there are further extensions possible.

The method itself can be extended by using additional algorithms for reference class selection.
Other methods for dimension reductions are possible, e.g., the self-organizing map \citep{kohonen82}, an artificial neural network using a two dimensional grid of neurons for dimension reduction.
On data sets with cluster structures, we could use the neurons of the self-organizing map as cluster centers or other cluster algorithms.
A parametric model for the distribution of outcomes within the reference classes \citep[as in][]{stanleyetal96} could be used for reference class forecasting.

Within our method, it is crucial to rank the forecast ability of the different algorithms and reference variables.
We have not answered whether the results on calibration differ statistically significantly between the forecasts and acceptable numerical regions of the accuracy measures for generating appropriate reference classes are unknown, yet.
The only indication so far is given by the results of the market climate approach that can be interpreted as a prediction of the marginal distribution but it would be quite useful to know which forecasts are in fact calibrated.
Given a data set with less missing values, we could then use scoring rules that additionally assess the sharpness of forecasts and are more suitable for comparative assessment \citep{gneitingrafterty2007}.

It is yet open, whether reference classes can identify underlying distributions which could be answered in a simulation study to deepen the unterstanding of the mechanism behind reference class selection.
A study on similarity based forecasting using a weighted mean of reference class outcomes to issue point forecasts is also possible.
On a similar line of thought, correcting potentially biased expert (or model based) forecasts with the outside views can be investigated as the original corrective procedure in \cite{kahnemantversky79} suggested.
This means that expert forecasts would be combined with a reference class forecasts and a backtest could check for forecast improvement.

Investigations on other data sets beyond the presented case are necessary to further advocate the utility of the approach.
Additional possible applications are characterized by availability of sufficient data on past outcomes and by the fact that forecasts should typically be hard to issue.
Ideally, no models producing calibrated (distributional) forecasts directly should be known in the literature or existing models should be very complicated and/or not accepted by a broad audience of practitioners and thus sparsely used.
In the field of finance, forecasting of cash flow items (in corporate value theory), bankruptcy probabilities (or credit rating) and financial returns to assess value at risk may be possible further applications.

\section*{Disclosure Statement}

The author reports there are no competing interests to declare.

\section*{Data Availability}

The ``CRSP daily stock'' and ``Compustat daily updates - fundamentals annual'' data that support the findings of this study are available from Wharton Research Data Services and were downloaded on 28 and 30 January, 2020, respectively. 
Restrictions apply to the availability of these data, which were used under license for this study. 
Data are available at https://wrds-www.wharton.upenn.edu/ with the permission of Wharton Research Data Services.
The consumer price index data that support the findings of this study are openly available at FRED (Federal Reserve
Economic Data) at https://fred.stlouisfed.org/series/CPIAUCSL/, reference CPIAUCSL, and were downloaded on 29 January, 2020.
The ``Core company data - estimates data'' that support the findings of this study are available from FactSet and were downloaded on 7 January, 2021.
Restrictions apply to the availability of these data, which were used under license for this study.
Data are available at http://factset.com/ with the permission of FactSet.

\bibliography{LiteratureReferenceClassSelection}

\appendix

\section{Supporting Tables}\label{app:refvartables}

This section contains tables with supporting results from the backtest in Section \ref{sec:backtest}.
Table \ref{tab:resrdcont} shows that more reference variables do not necessarily improve reference class selection based on rank deviation (c.f. the discussion in Section \ref{subsec:resultsbacktest}).
Tables \ref{tab:resh1forward}, \ref{tab:resh3forward}, \ref{tab:resh5forward} and \ref{tab:resh10forward} contain the best results from the forward selection of the rank deviation procedure for all considered forecast horizons as described in Section \ref{subsec:varselection}.
The best three sets of four, five and six reference variables from the forward selection are used in backtesting PCA rank deviation, see Section \ref{subsec:varselection}.
Tables \ref{tab:resh1brute}, \ref{tab:resh3brute}, \ref{tab:resh5brute} and \ref{tab:resh10brute} contain the best results from the brute force backtest of rank deviation on contemporaneous reference variables for all considered forecast horizons as described in Section \ref{subsec:varselection}.
The best three sets of four, five and six reference variables from the brute force approach are used in the backtest of PCA rank deviation as well, see Section \ref{subsec:varselection}.

\begin{table}[H]
\center
\caption{Best results for combining contemporaneous reference variables with past sales growth rates and past operating margin deltas up to different lags for horizons one, three, five and ten.}
\label{tab:resrdcont}
\begin{tabular}{llllll|rrr}
  Horizon & Ref.Var. & Comb. & Cor. & $w$ & Size & $\Delta_{\text{q}}$ & KS & CvM \\ 
  \midrule
  \multirow{3}{*}{1} & contemp., & union & no & 30 & 0.05 & 0.0584 & 7.0061 & 15.9007 \\ 
  &  $\text{salesGR}_{1}$, &  &  &  &  &  &  &  \\
  &  $\text{opmar}\Delta_{1}$ &  &  &  &  &  &  &  \\
  \midrule
  \multirow{3}{*}{3} & contemp., & union & no & 30 & 0.01 & 0.0773 & 8.0135 & 24.0522 \\
  &  $\text{salesGR}_{1}$, &  &  &  &  &  &  &  \\ 
  &  $\text{opmar}\Delta_{1}$ &  &  &  &  &  &  &  \\
  \midrule
  \multirow{3}{*}{5} & contemp., & union & yes & 30 & 0.01 & 0.0886 & 8.4625 & 26.1944 \\ 
  &  $\text{salesGR}_{1}$, &  &  &  &  &  &  &  \\
  &  $\text{opmar}\Delta_{1}$ &  &  &  &  &  &  &  \\
  \midrule
  \multirow{3}{*}{10} & contemp., & union & yes & 30 & 0.01 & 0.0758 & 4.7809 & 6.7344 \\ 
  &  $\text{salesGR}_{1:5}$, &  &  &  &  &  &  &  \\
  &  $\text{opmar}\Delta_{1:5}$ &  &  &  &  &  &  &  \\
\end{tabular}
\end{table}

\begin{table}[H]
\center
\caption{Forward selection results for forecasting 1-year sales growth. The three best sets of four, five and six reference variable, respectively, are used in the backtest of PCA rank deviation.}
\label{tab:resh1forward}
\begin{tabular}{llllll|rrr}
  Choice & Ref.Var. & Comb. & Cor. & $w$ & Size & $\Delta_{\text{q}}$ & KS & CvM \\ 
  \midrule
  \multirow{6}{*}{Best single} & $\text{opmar}\Delta_{6}$ & -- & -- & 30 & 0.025 & 0.0157 & 1.8644 & 0.8256 \\ 
   & $\text{opmar}\Delta_{7}$ & -- & -- & 30 & 0.025 & 0.0159 & 2.2179 & 1.0808 \\ 
   & $\text{opmar}\Delta_{5}$ & -- & -- & 30 & 0.01 & 0.0171 & 2.3873 & 1.2154 \\ 
   & $\text{opmar}\Delta_{9}$ & -- & -- & 30 & 0.025 & 0.0187 & 2.4281 & 1.1636 \\ 
   & $\text{opmar}\Delta_{10}$ & -- & -- & 30 & 0.01 & 0.0188 & 2.0381 & 0.7830 \\ 
   & $\text{opmar}\Delta_{3}$ & -- & -- & 30 & 0.01 & 0.0202 & 2.5357 & 1.6515 \\ 
  \midrule
  \multirow{6}{*}{Best 2} & $\text{salesGR}_{7}$,  & union & no & 30 & 0.05 & 0.0114 & 1.6476 & 0.5886 \\ 
  & $\text{opmar}\Delta_{5}$  &  &  &  &  &  &  &  \\
   & $\text{salesGR}_{8}$,  & union & no & 30 & 0.05 & 0.0114 & 1.5929 & 0.5830 \\ 
  & $\text{opmar}\Delta_{6}$  &  &  &  &  &  &  &  \\
   & $\text{salesGR}_{5}$,  & union & yes & 30 & 0.05 & 0.0115 & 0.9091 & 0.1290 \\ 
  & $\text{opmar}\Delta_{7}$  &  &  &  &  &  &  &  \\
  \midrule
  \multirow{6}{*}{Best 3} & $\text{salesGR}_{5, 6}$, & union & yes & 30 & 0.01 & 0.0087 & 0.8496 & 0.1021 \\ 
  & $\text{opmar}\Delta_{7}$  &  &  &  &  &  &  &  \\
   & $\text{salesGR}_{5, 7}$, & union & no & 30 & 0.025 & 0.0098 & 0.8259 & 0.0843 \\ 
  & $\text{opmar}\Delta_{5}$  &  &  &  &  &  &  &  \\
   & $\text{salesGR}_{5, 7}$, & union & yes & 30 & 0.01 & 0.0098 & 0.9199 & 0.1943 \\
  & $\text{opmar}\Delta_{7}$  &  &  &  &  &  &  &  \\
  \midrule
  \multirow{6}{*}{Best 4} & $\text{salesGR}_{5:7}$, & union & yes & 30 & 0.01 & 0.0066 & 0.7408 & 0.0448 \\ 
  & $\text{opmar}\Delta_{5}$  &  &  &  &  &  &  &  \\
   & $\text{salesGR}_{5, 7, 8}$, & union & no & 30 & 0 & 0.0072 & 0.6538 & 0.0717 \\ 
  & $\text{opmar}\Delta_{5}$  &  &  &  &  &  &  &  \\
   & $\text{salesGR}_{3, 5, 6}$, & union & no & 30 & 0 & 0.0086 & 0.8131 & 0.1163 \\ 
  & $\text{opmar}\Delta_{7}$  &  &  &  &  &  &  &  \\ 
  \midrule
  \multirow{6}{*}{Best 5} & $\text{salesGR}_{3, 5:7}$, & union & no & 30 & 0.05 & 0.0065 & 0.7191 & 0.0407 \\ 
  & $\text{opmar}\Delta_{5}$  &  &  &  &  &  &  &  \\
   & $\text{salesGR}_{5:8}$, & union & no & 30 & 0.05 & 0.0072 & 0.6475 & 0.0698 \\ 
  & $\text{opmar}\Delta_{5}$  &  &  &  &  &  &  &  \\
   & $\text{salesGR}_{3:6}$, & union & no & 30 & 0.05 & 0.0084 & 0.6955 & 0.0546 \\
  & $\text{opmar}\Delta_{7}$  &  &  &  &  &  &  &  \\
  \midrule 
  \multirow{6}{*}{Best 6} & $\text{salesGR}_{3:7}$, & union & no & 30 & 0.05 & 0.0076 & 0.7212 & 0.0628 \\ 
  & $\text{opmar}\Delta_{5}$  &  &  &  &  &  &  &  \\
   & $\text{salesGR}_{3:7}$, & union & no & 30 & 0.025 & 0.0093 & 0.7668 & 0.1116 \\ 
  & $\text{opmar}\Delta_{7}$  &  &  &  &  &  &  &  \\
   & $\text{salesGR}_{5:8}$, & union & no & 30 & 0.05 & 0.0093 & 0.8306 & 0.1569 \\ 
  & $\text{opmar}\Delta_{5, 6}$  &  &  &  &  &  &  &  \\
\end{tabular}
\end{table}

\begin{table}[H]
\center
\caption{Brute force results for forecasting 1-year sales growth. The three best sets of four, five and six reference variable, respectively, are used in the backtest of PCA rank deviation.}
\label{tab:resh1brute}
\begin{tabular}{llllll|rrr}
  Choice & Ref.Var. & Comb. & Cor. & $w$ & Size & $\Delta_{\text{q}}$ & KS & CvM \\ 
  \midrule
  \multirow{3}{*}{Best} & $\beta$, P/E & union & yes & 5 & 0.05 & 0.0158 & 2.7215 & 1.8060 \\ 
  & $\beta$, P/E, P/B & union & no & 5 & 0.05 & 0.0199 & 2.4706 & 1.5982 \\ 
  & P/E, P/B & union & no & 5 & 0.05 & 0.0251 & 3.0271 & 2.8965 \\ 
  \midrule
  \multirow{3}{*}{Best 4} & at, $\beta$, P/E, P/B & union & no & 5 & 0.05 & 0.0306 & 4.3259 & 4.4872 \\ 
  & seq, $\beta$, P/E, P/B & union & no & 5 & 0.05 & 0.0316 & 4.1763 & 4.4604 \\ 
  & at, seq, $\beta$, P/E & union & no & 5 & 0.05 & 0.0320 & 5.0380 & 6.2068 \\ 
  \midrule
  \multirow{3}{*}{Best 5} & at, seq, $\beta$, P/E, P/B & union & no & 5 & 0.05 & 0.0323 & 4.6028 & 5.0844 \\ 
  & sales, at, seq, P/E, P/B & union & no & 5 & 0.05 & 0.0374 & 5.0118 & 6.1482 \\ 
  & sales, at, $\beta$, P/E, P/B & union & yes & 5 & 0.01 & 0.0379 & 5.4581 & 7.0826 \\ 
  \midrule
  \multirow{6}{*}{Best 6} & sales, at, seq, $\beta$ & union & no & 5 & 0.05 & 0.0377 & 4.9120 & 6.0955 \\ 
  &  P/E, P/B &  &  &  &  &  &  &  \\
  & sales, opmar, at, $\beta$, & union & yes & 5 & 0.01 & 0.0437 & 5.2103 & 7.8767 \\ 
  &  P/E, P/B &  &  &  &  &  &  &  \\
  & sales, opmar, at, seq, & union & yes & 5 & 0.01 & 0.0447 & 5.2960 & 8.0278 \\ 
  &  P/E, P/B &  &  &  &  &  &  &  \\
\end{tabular}
\end{table}

\begin{table}[H]
\center
\caption{Forward selection results for forecasting 3-year sales growth. The three best sets of four, five and six reference variable, respectively, are used in the backtest of PCA rank deviation.}
\label{tab:resh3forward}
\begin{tabular}{llllll|rrr}
  Choice & Ref.Var. & Comb. & Cor. & $w$ & Size & $\Delta_{\text{q}}$ & KS & CvM \\ 
  \midrule
	\multirow{6}{*}{Best single} & $\text{opmar}\Delta_{6}$ & -- & -- & 30 & 0.025 & 0.0157 & 1.8644 & 0.8256 \\ 
   & $\text{opmar}\Delta_{8}$ & -- & -- & 30 & 0.025 & 0.0296 & 2.0007 & 1.0991 \\ 
   & $\text{opmar}\Delta_{10}$ & -- & -- & 30 & 0.01 & 0.0319 & 2.0996 & 1.1686 \\ 
   & $\text{opmar}\Delta_{7}$ & -- & -- & 30 & 0.01 & 0.0321 & 3.4675 & 3.4485 \\ 
   & $\text{opmar}\Delta_{9}$ & -- & -- & 30 & 0.01 & 0.0348 & 2.1389 & 1.1537 \\ 
   & $\text{opmar}\Delta_{5}$ & -- & -- & 30 & 0.01 & 0.0363 & 5.5770 & 8.9425 \\ 
  \midrule
  \multirow{5}{*}{Best 2} & $\text{salesGR}_{10}$, & union & yes & 30 & 0.01 & 0.0239 & 1.4898 & 0.5776 \\ 
  	& $\text{opmar}\Delta_{8}$ &  &  &  &  &  &  &  \\
   & opmar, & inters. & no & 30 & 0.05 & 0.0242 & 1.8007 & 0.7059 \\ 
  	& $\text{opmar}\Delta_{7}$ &  &  &  &  &  &  &  \\
   & $\text{opmar}\Delta_{8, 10}$ & union & no & 30 & 0.05 & 0.0260 & 2.0310 & 1.0684 \\ 
  \midrule
  \multirow{5}{*}{Best 3} & $\text{salesGR}_{10}$, & union & yes & 30 & 0.01 & 0.0223 & 1.7135 & 0.7237 \\
  	& $\text{opmar}\Delta_{8, 9}$ &  &  &  &  &  &  &  \\ 
   & $\text{salesGR}_{10}$, & union & yes & 30 & 0.01 & 0.0255 & 1.7001 & 0.6329 \\ 
  	& $\text{opmar}\Delta_{8, 10}$ &  &  &  &  &  &  &  \\
   & $\text{opmar}\Delta_{8:10}$ & union & no & 30 & 0.05 & 0.0261 & 1.9815 & 0.9914 \\ 
  \midrule
  \multirow{6}{*}{Best 4} & $\text{salesGR}_{10}$, & union & yes & 30 & 0.01 & 0.0233 & 1.7000 & 0.6886 \\ 
  	& $\text{opmar}\Delta_{8:10}$ &  &  &  &  &  &  &  \\
   & $\text{salesGR}_{9}$, & union & yes & 30 & 0.01 & 0.0257 & 1.8971 & 0.7147 \\ 
  	& $\text{opmar}\Delta_{8:10}$ &  &  &  &  &  &  &  \\
   & $\text{salesGR}_{8}$, & union & no & 30 & 0.01 & 0.0269 & 1.7087 & 0.7064 \\ 
  	& $\text{opmar}\Delta_{8:10}$ &  &  &  &  &  &  &  \\
  \midrule
  \multirow{6}{*}{Best 5} & $\text{salesGR}_{9, 10}$, & union & yes & 30 & 0.01 & 0.0268 & 1.6305 & 0.6640 \\ 
  	& $\text{opmar}\Delta_{8:10}$ &  &  &  &  &  &  &  \\
   & $\text{salesGR}_{8, 9}$, & union & yes & 30 & 0.01 & 0.0279 & 1.7261 & 0.6208 \\ 
  	& $\text{opmar}\Delta_{8:10}$ &  &  &  &  &  &  &  \\
   & $\text{salesGR}_{8, 10}$, & union & yes & 30 & 0.025 & 0.0283 & 1.6882 & 0.7032 \\ 
  	& $\text{opmar}\Delta_{8:10}$ &  &  &  &  &  &  &  \\
  \midrule
  \multirow{6}{*}{Best 6} & $\text{salesGR}_{8:10}$, & union & yes & 30 & 0.01 & 0.0293 & 1.7441 & 0.8076 \\ 
  	& $\text{opmar}\Delta_{8:10}$ &  &  &  &  &  &  &  \\
   & $\text{salesGR}_{7, 9, 10}$, & union & yes & 30 & 0.01 & 0.0299 & 2.5688 & 1.7311 \\ 
  	& $\text{opmar}\Delta_{8:10}$ &  &  &  &  &  &  &  \\
   & $\text{salesGR}_{7, 8, 10}$, & union & yes & 30 & 0.01 & 0.0306 & 2.5785 & 1.8136 \\ 
  	& $\text{opmar}\Delta_{8:10}$ &  &  &  &  &  &  &  \\
\end{tabular}
\end{table}

\begin{table}[H]
\center
\caption{Brute force results for forecasting 3-year sales growth. The three best sets of four, five and six reference variable, respectively, are used in the backtest of PCA rank deviation.}
\label{tab:resh3brute}
\begin{tabular}{llllll|rrr}
  Choice & Ref.Var. & Comb. & Cor. & $w$ & Size & $\Delta_{\text{q}}$ & KS & CvM \\ 
  \midrule
	\multirow{3}{*}{Best} & $\beta$, P/E, P/B & LARD & -- & 10 & 0.05 & 0.0319 & 4.7188 & 6.6191 \\ 
  	  & $\beta$, P/E & LARD & -- & 30 & 0.05 & 0.0388 & 4.5036 & 6.1123 \\ 
    & opmar, $\beta$ & union & yes & 30 & 0.01 & 0.0392 & 4.7713 & 7.8706 \\ 
  	\midrule
  	\multirow{3}{*}{Best 4} & at, $\beta$, P/E, P/B & union & no & 5 & 0.05 & 0.0594 & 9.4340 & 23.8707 \\ 
    & sales, $\beta$, P/E, P/B & union & no & 5 & 0.05 & 0.0655 & 9.1517 & 21.6102 \\ 
    & sales, opmar, $\beta$, P/E & union & yes & 30 & 0.01 & 0.0670 & 7.0775 & 18.7256 \\ 
  	\midrule
  	\multirow{5}{*}{Best 5} & at, seq, $\beta$, P/E, P/B & union & no & 5 & 0.05 & 0.0680 & 9.9804 & 28.7600 \\ 
   	& sales, at, $\beta$, & union & no & 5 & 0.05 & 0.0688 & 10.1119 & 27.1444 \\ 
   	&  P/E, P/B &  &  &  &  &  &  &  \\
   	& sales, opmar, $\beta$, & union & no & 5 & 0.05 & 0.0738 & 9.3260 & 24.5716 \\ 
    &  P/E, P/B &  &  &  &  &  &  &  \\
  	\midrule
  	\multirow{6}{*}{Best 6} & sales, at, seq, $\beta$, & union & no & 5 & 0.05 & 0.0753 & 10.4470 & 30.4821 \\ 
  	&  P/E, P/B &  &  &  &  &  &  &  \\
  	& sales, opmar, at, $\beta$, & union & no & 5 & 0.05 & 0.0769 & 10.2567 & 29.2024 \\ 
  	&  P/E, P/B &  &  &  &  &  &  &  \\
  	& sales, opmar, seq, $\beta$, & union & yes & 30 & 0.01 & 0.0816 & 8.8320 & 28.1591 \\ 
  	&  P/E, P/B &  &  &  &  &  &  &  \\
\end{tabular}
\end{table}

\begin{table}[H]
\center
\caption{Forward selection results for forecasting 5-year sales growth. The three best sets of four, five and six reference variable, respectively, are used in the backtest of PCA rank deviation.}
\label{tab:resh5forward}
\begin{tabular}{llllll|rrr}
  Choice & Ref.Var. & Comb. & Cor. & $w$ & Size & $\Delta_{\text{q}}$ & KS & CvM \\ 
  \midrule
	\multirow{6}{*}{Best single} & $\text{opmar}\Delta_{6}$ & -- & -- & 30 & 0.025 & 0.0157 & 1.8644 & 0.8256 \\ 
   & $\text{opmar}\Delta_{10}$ & -- & -- & 30 & 0.025 & 0.0337 & 1.7525 & 0.9926 \\ 
   & $\text{opmar}\Delta_{5}$ & -- & -- & 30 & 0.01 & 0.0371 & 3.8690 & 4.0986 \\ 
   & $\text{opmar}\Delta_{9}$ & -- & -- & 30 & 0.01 & 0.0375 & 2.0446 & 1.4290 \\ 
   & $\text{opmar}\Delta_{7}$ & -- & -- & 30 & 0.01 & 0.0409 & 2.5724 & 1.7135 \\ 
   & $\text{opmar}\Delta_{4}$ & -- & -- & 30 & 0.01 & 0.0417 & 4.9117 & 7.5028 \\ 
  \midrule
  \multirow{6}{*}{Best 2} & $\text{salesGR}_{10}$, & union & yes & 30 & 0.01 & 0.0230 & 2.2279 & 1.0487 \\ 
  	& $\text{opmar}\Delta_{6}$ &  &  &  &  &  &  &  \\
   & $\text{salesGR}_{9}$, & union & yes & 30 & 0.01 & 0.0271 & 1.9119 & 0.9894 \\ 
  	& $\text{opmar}\Delta_{6}$ &  &  &  &  &  &  &  \\
   & $\text{salesGR}_{10}$,& union & yes & 30 & 0.01 & 0.0272 & 1.7697 & 1.0131 \\ 
  	& $\text{opmar}\Delta_{10}$ &  &  &  &  &  &  &  \\
  \midrule
  \multirow{6}{*}{Best 3} & $\text{salesGR}_{6, 10}$, & union & yes & 30 & 0.01 & 0.0264 & 1.5280 & 0.6581 \\ 
  	& $\text{opmar}\Delta_{6}$ &  &  &  &  &  &  &  \\
   & $\text{salesGR}_{8, 10}$, & union & yes & 30 & 0.01 & 0.0265 & 1.7605 & 0.7962 \\ 
  	& $\text{opmar}\Delta_{6}$ &  &  &  &  &  &  &  \\
   & $\text{salesGR}_{7, 10}$, & union & yes & 30 & 0.01 & 0.0275 & 1.5368 & 0.6862 \\ 
  	& $\text{opmar}\Delta_{6}$ &  &  &  &  &  &  &  \\
  \midrule
  \multirow{6}{*}{Best 4} & $\text{salesGR}_{5, 8, 10}$, & union & yes & 30 & 0.01 & 0.0265 & 2.3651 & 1.4054 \\ 
  	& $\text{opmar}\Delta_{6}$ &  &  &  &  &  &  &  \\
   & $\text{salesGR}_{5, 6, 10}$, & union & yes & 30 & 0.05 & 0.0267 & 2.4667 & 1.6128 \\ 
  	& $\text{opmar}\Delta_{6}$ &  &  &  &  &  &  &  \\
   & $\text{salesGR}_{5, 7, 10}$, & union & yes & 30 & 0.01 & 0.0267 & 2.5496 & 1.5178 \\ 
  	& $\text{opmar}\Delta_{6}$ &  &  &  &  &  &  &  \\
  \midrule
  \multirow{6}{*}{Best 5} & $\text{salesGR}_{5:7, 10}$, & union & yes & 30 & 0.05 & 0.0271 & 2.3006 & 1.3987 \\
  	& $\text{opmar}\Delta_{6}$ &  &  &  &  &  &  &  \\ 
   & $\text{salesGR}_{5, 6, 8, 10}$, & union & yes & 30 & 0.025 & 0.0271 & 2.1439 & 1.2915 \\ 
  	& $\text{opmar}\Delta_{6}$ &  &  &  &  &  &  &  \\
   & $\text{salesGR}_{5, 6, 9, 10}$, & union & yes & 30 & 0.01 & 0.0279 & 2.1598 & 1.2389 \\ 
  	& $\text{opmar}\Delta_{6}$ &  &  &  &  &  &  &  \\
  \midrule
  \multirow{6}{*}{Best 6} & $\text{salesGR}_{5, 6, 9, 10}$, & union & yes & 30 & 0.01 & 0.0275 & 2.5795 & 1.5061 \\ 
  	& $\text{opmar}\Delta_{6, 8}$ &  &  &  &  &  &  &  \\
   & $\text{salesGR}_{5:7, 9, 10}$, & union & yes & 30 & 0.01 & 0.0277 & 1.9939 & 1.2312 \\ 
  	& $\text{opmar}\Delta_{6}$ &  &  &  &  &  &  &  \\
   & $\text{salesGR}_{5:7, 10}$, & union & yes & 30 & 0.01 & 0.0279 & 2.5866 & 1.6759 \\ 
  	& $\text{opmar}\Delta_{6, 8}$ &  &  &  &  &  &  &  \\
\end{tabular}
\end{table}

\begin{table}[H]
\center
\caption{Brute force results for forecasting 5-year sales growth. The three best sets of four, five and six reference variable, respectively, are used in the backtest of PCA rank deviation.}
\label{tab:resh5brute}
\begin{tabular}{llllll|rrr}
  Choice & Ref.Var. & Comb. & Cor. & $w$ & Size & $\Delta_{\text{q}}$ & KS & CvM \\ 
  \midrule
	\multirow{3}{*}{Best} & $\beta$, P/E & LARD & -- & 30 & 0.05 & 0.0492 & 5.1695 & 6.0082 \\ 
  	& opmar, $\beta$ & union & yes & 30 & 0.01 & 0.0525 & 5.7728 & 10.7982 \\ 
  	& opmar, seq, P/B & inters. & no & 30 & 0.025 & 0.0543 & 0.6534 & 0.0646 \\ 
  	\midrule
  	\multirow{6}{*}{Best 4}  & sales, opmar, & union & yes & 30 & 0.025 & 0.0781 & 7.6348 & 21.5909 \\ 
  	&  $\beta$, P/E &  &  &  &  &  &  &  \\
  	& sales, opmar, & union & yes & 30 & 0.01 & 0.0864 & 8.0867 & 26.1398 \\ 
  	& seq, P/E &  &  &  &  &  &  &  \\
  	& sales, opmar, & union & yes & 30 & 0.01 & 0.0871 & 9.0112 & 25.3800 \\ 
  	&  $\beta$, P/B &  &  &  &  &  &  &  \\
  	\midrule
  	\multirow{6}{*}{Best 5}  & sales, opmar, $\beta$, & union & yes & 30 & 0.01 & 0.0894 & 8.8936 & 25.6606 \\ 
  	&  P/E, P/B &  &  &  &  &  &  &  \\
  	& sales, opmar, at, & union & yes & 30 & 0.01 & 0.0931 & 8.5649 & 29.8700 \\ 
  	&  $\beta$, P/E &  &  &  &  &  &  &  \\
  	& sales, opmar, seq, & union & yes & 30 & 0.01 & 0.0943 & 8.2488 & 29.5979 \\ 
  	&  $\beta$, P/E &  &  &  &  &  &  &  \\
  	\midrule
  	\multirow{6}{*}{Best 6}  & sales, opmar, seq, & union & yes & 30 & 0.01 & 0.0975 & 9.0194 & 30.8282 \\ 
  	& $\beta$, P/E, P/B &  &  &  &  &  &  &  \\
 	& sales, opmar, at, & union & yes & 30 & 0.01 & 0.1000 & 9.3992 & 31.4917 \\
  	&  $\beta$, P/E, P/B &  &  &  &  &  &  &  \\ 
  	& sales, opmar, at, & union & yes & 30 & 0.01 & 0.1045 & 9.0507 & 36.2843 \\ 
  	&  seq, $\beta$, P/E &  &  &  &  &  &  &  \\
\end{tabular}
\end{table}

\begin{table}[H]
\center
\caption{Forward selection results for forecasting 10-year sales growth. The three best sets of four, five and six reference variable, respectively, are used in the backtest of PCA rank deviation.}
\label{tab:resh10forward}
\begin{tabular}{llllll|rrr}
  Choice & Ref.Var. & Comb. & Cor. & $w$ & Size & $\Delta_{\text{q}}$ & KS & CvM \\ 
  \midrule
	\multirow{6}{*}{Best single} & $\text{opmar}\Delta_{6}$ & -- & -- & 30 & 0.025 & 0.0157 & 1.8644 & 0.8256 \\ 
   & $\text{opmar}\Delta_{7}$ & -- & -- & 30 & 0.025 & 0.0461 & 3.5950 & 3.8367 \\ 
   & $\text{opmar}\Delta_{5}$ & -- & -- & 30 & 0.025 & 0.0482 & 4.1254 & 5.0874 \\ 
   & $\text{opmar}\Delta_{4}$ & -- & -- & 30 & 0.01 & 0.0510 & 4.7700 & 6.4086 \\ 
   & $\text{opmar}\Delta_{8}$ & -- & -- & 30 & 0.025 & 0.0520 & 3.8959 & 4.4523 \\ 
   & $\text{opmar}\Delta_{9}$ & -- & -- & 30 & 0.05 & 0.0585 & 3.8342 & 4.8646 \\ 
  \midrule
	\multirow{5}{*}{Best 2} & $\text{salesGR}_{6}$, & union & yes & 30 & 0.01 & 0.0401 & 3.1193 & 3.2598 \\ 
  	& $\text{opmar}\Delta_{7}$ &  &  &  &  &  &  &  \\
   & $\text{opmar}\Delta_{5, 6}$ & union & yes & 30 & 0.05 & 0.0415 & 4.1217 & 4.9332 \\ 
   & $\text{salesGR}_{5}$, & union & yes & 30 & 0.01 & 0.0424 & 3.2929 & 3.3610 \\ 
  	& $\text{opmar}\Delta_{7}$ &  &  &  &  &  &  &  \\
  \midrule
	\multirow{6}{*}{Best 3} & $\text{salesGR}_{6}$, & union & yes & 30 & 0.05 & 0.0409 & 3.5038 & 3.7480 \\ 
  	& $\text{opmar}\Delta_{5, 7}$ &  &  &  &  &  &  &  \\
   & $\text{salesGR}_{7}$, & union & yes & 30 & 0.05 & 0.0410 & 3.5776 & 3.9702 \\ 
  	& $\text{opmar}\Delta_{5, 6}$ &  &  &  &  &  &  &  \\
   & $\text{salesGR}_{6}$, & union & yes & 30 & 0.01 & 0.0416 & 3.6444 & 3.8220 \\ 
  	& $\text{opmar}\Delta_{5, 6}$ &  &  &  &  &  &  &  \\
  \midrule
	\multirow{6}{*}{Best 4} & $\text{salesGR}_{6, 7}$, & union & yes & 30 & 0.01 & 0.0388 & 3.0820 & 3.2019 \\ 
  	& $\text{opmar}\Delta_{5, 7}$ &  &  &  &  &  &  &  \\
   & $\text{salesGR}_{7}$, & union & yes & 30 & 0.05 & 0.0411 & 3.4968 & 3.5954 \\ 
  	& $\text{opmar}\Delta_{5:7}$ &  &  &  &  &  &  &  \\
   & $\text{salesGR}_{5, 6}$, & union & yes & 30 & 0.01 & 0.0423 & 3.4630 & 3.6062 \\ 
  	& $\text{opmar}\Delta_{5, 6}$ &  &  &  &  &  &  &  \\
  \midrule
	\multirow{6}{*}{Best 5} & $\text{salesGR}_{5:7}$, & union & yes & 30 & 0.025 & 0.0419 & 3.2042 & 3.3900 \\ 
  	& $\text{opmar}\Delta_{5, 7}$ &  &  &  &  &  &  &  \\
   & $\text{salesGR}_{6:8}$, & union & yes & 30 & 0.01 & 0.0422 & 3.0932 & 3.3684 \\ 
  	& $\text{opmar}\Delta_{5, 7}$ &  &  &  &  &  &  &  \\
   & $\text{salesGR}_{5, 6}$, & union & yes & 30 & 0.01 & 0.0425 & 3.7850 & 4.3952 \\ 
  	& $\text{opmar}\Delta_{4:6}$ &  &  &  &  &  &  &  \\
  \midrule
	\multirow{6}{*}{Best 6} & $\text{salesGR}_{6:8}$, & union & yes & 30 & 0.01 & 0.0403 & 2.9479 & 2.9308 \\ 
  	& $\text{opmar}\Delta_{5, 7, 8}$ &  &  &  &  &  &  &  \\
   & $\text{salesGR}_{5:7}$, & union & yes & 30 & 0.025 & 0.0420 & 3.4675 & 3.5151 \\ 
  	& $\text{opmar}\Delta_{5:7}$ &  &  &  &  &  &  &  \\
   & $\text{salesGR}_{6:9}$, & union & yes & 30 & 0.01 & 0.0432 & 3.0288 & 3.4022 \\ 
  	& $\text{opmar}\Delta_{5, 7}$ &  &  &  &  &  &  &  \\
\end{tabular}
\end{table}

\begin{table}[H]
\center
\caption{Brute force results for forecasting 10-year sales growth. The three best sets of four, five and six reference variable, respectively, are used in the backtest of PCA rank deviation.}
\label{tab:resh10brute}
\begin{tabular}{llllll|rrr}
  	Choice & Ref.Var. & Comb. & Cor. & $w$ & Size & $\Delta_{\text{q}}$ & KS & CvM \\ 
  	\midrule
	\multirow{3}{*}{Best} & $\beta$, P/E & LARD & -- & 30 & 0.025 & 0.0680 & 4.1868 & 4.2020 \\ 
  	 & opmar, P/E & union & yes & 30 & 0.01 & 0.0698 & 4.5438 & 7.8400 \\ 
  	 & at, $\beta$, P/E, P/B & LARD & -- & 30 & 0.05 & 0.0774 & 6.1698 & 10.9065 \\
  	\midrule 
  	\multirow{4}{*}{Best 4} & opmar, $\beta$, P/E, & LARD & -- & 20 & 0.025 & 0.0804 & 9.2137 & 19.5756 \\ 
  	 &  P/B  &  &  &  &  &  &  &  \\
  	 & opmar, at, $\beta$, P/E & LARD & -- & 20 & 0.05 & 0.0941 & 11.6692 & 31.5442 \\ 
  	 & sales, $\beta$, P/E, P/B & LARD & -- & 30 & 0.05 & 0.0956 & 7.3187 & 16.8942 \\ 
  	\midrule
  	\multirow{6}{*}{Best 5} & opmar, at, $\beta$, & LARD & -- & 20 & 0.05 & 0.0989 & 8.7529 & 21.4948 \\ 
  	&  P/E, P/B  &  &  &  &  &  &  &  \\
  	 & sales, opmar, at, & inters. & yes & 30 & 0.05 & 0.1059 & 4.1239 & 6.7161 \\ 
  	&   seq, $\beta$ &  &  &  &  &  &  &  \\
  	 & sales, opmar, $\beta$, & union & yes & 30 & 0.01 & 0.1217 & 8.2634 & 22.2247 \\ 
  	&   P/E, P/B &  &  &  &  &  &  &  \\
  	\midrule
  	\multirow{6}{*}{Best 6} & sales, opmar, at,  & union & yes & 30 & 0.01 & 0.1389 & 9.1815 & 26.8634 \\ 
  	&  $\beta$, P/E, P/B &  &  &  &  &  &  &  \\
  	 & sales, opmar, seq,  & union & yes & 30 & 0.025 & 0.1414 & 9.7186 & 28.1473 \\ 
  	&  $\beta$, P/E, P/B &  &  &  &  &  &  &  \\
  	 & sales, opmar, at, & union & yes & 30 & 0.01 & 0.1443 & 9.2700 & 28.0870 \\ 
  	&  seq, P/E, P/B  &  &  &  &  &  &  &  \\
\end{tabular}
\end{table}

\end{document}